\documentclass[a4paper,11pt]{article}
\usepackage{amssymb}
\usepackage{amsmath}
\usepackage{latexsym,amssymb,amsmath}
\usepackage{graphicx}
\usepackage{epsfig}

\DeclareFontFamily{U}{rsf}{} \DeclareFontShape{U}{rsf}{m}{n}{
  <5> <6> rsfs5 <7> <8> <9> rsfs7 <10-> rsfs10}{}
\DeclareMathAlphabet\Scr{U}{rsf}{m}{n} \makeatletter
\@addtoreset{equation}{section} \makeatother

\newcommand{\be}{\begin{equation}}
\newcommand{\ee}{\end{equation}}
\newcommand{\bea}{\begin{eqnarray}}
\newcommand{\eea}{\end{eqnarray}}
\newcommand{\ba}{\begin{array}}
\newcommand{\ea}{\end{array}}
\newcommand{\bit}{\begin{itemize}}
\newcommand{\eit}{\end{itemize}}
\newcommand{\ben}{\begin{enumerate}}
\newcommand{\een}{\end{enumerate}}

\begin{document}

\begin{titlepage}
 \thispagestyle{empty}
\begin{flushright}
     \hfill{CERN-PH-TH/2010-272}\\
 \end{flushright}

 \vspace{50pt}

 \begin{center}
     { \huge{\bf      {Two-Center Black Holes Duality-Invariants \\
     \vspace{5pt}for $stu$ Model and its lower-rank Descendants
    }}}

     \vspace{25pt}

     {\Large {Sergio Ferrara$^{a,b,c}$, Alessio Marrani$^{a}$, Emanuele Orazi$^{a}$ \\ Raymond Stora$^{a,d}$ and Armen Yeranyan$^{b,e}$}}

     \vspace{20pt}

  {\it ${}^a$ Physics Department, Theory Unit, CERN,\\
     CH -1211, Geneva 23, Switzerland;\\
     \texttt{sergio.ferrara@cern.ch}\\
     \texttt{marrani@lnf.infn.it}\\
     \texttt{orazi@lnf.infn.it}}

     \vspace{10pt}

    {\it ${}^b$ INFN - Laboratori Nazionali di Frascati,\\
     Via Enrico Fermi 40, I-00044 Frascati, Italy\\
     \texttt{ayeran@lnf.infn.it}}

     \vspace{10pt}

     {\it ${}^c$  Department of Physics and Astronomy,\\
University of California, Los Angeles, CA 90095-1547,USA}\\

     \vspace{10pt}

     {\it ${}^d$  LAPTH, Universit\'{e} de Savoie, CNRS\\
Annecy-le-Vieux, France}\\

     \vspace{10pt}

{\it ${}^e$ Department of Physics, Yerevan State University\\
Alex Manoogian St. 1, Yerevan, 0025, Armenia}\\

     \vspace{10pt}

     \vspace{30pt}

     {ABSTRACT}

 \vspace{10pt}
 \end{center}
We classify $2$-center extremal black hole charge configurations
through duality-invariant homogeneous polynomials, which are the
generalization of the unique
invariant quartic polynomial for single-center black holes based on homogeneous symmetric cubic special K\"{a}%
hler geometries.

A crucial role is played by an horizontal $SL(p,\mathbb{R})$
symmetry group, which classifies invariants for $p$-center black
holes. For $p=2$, a (spin $2 $) quintet of quartic invariants
emerge. We provide the minimal set of independent invariants for the
rank-$3$ $\mathcal{N}=2$, $d=4$ $stu$ model, and for its lower-rank
descendants, namely the rank-$2$ $st^{2}$ and rank-$1$ $t^{3}$
models; these models respectively exhibit seven, six and five
independent invariants.

We also derive the polynomial relations among these and other
duality invariants. In particular, the symplectic product of two
charge vectors is not independent from the quartic quintet in the
$t^{3}$ model, but rather it satisfies a degree-$16$ relation,
corresponding to a quartic equation for the square of the symplectic
product itself.
\end{titlepage}
\tableofcontents
\section{\label{Intro}Introduction}

Multi-center black holes (BHs) are a natural extension of single-center BHs,
and they play an important role in the dynamics of quantum theories of
gravity, such as superstrings and $M$-theory.

In fact, interesting multi-center solutions have been found for BPS BHs in $%
d=4$ theories with $\mathcal{N}=2$ supersymmetry, in which the \textit{%
Attractor Mechanism} \cite{AM-Refs,FGK} for static, spherically symmetric,
asymptotically flat, extremal dyonic BHs is generalised by the so-called
\textit{split attractor flow} \cite{D-1,D-2,BD-1}. This name comes from the
existence, for $2$-center solutions, of a co-dimension one region (named
\textit{marginal stability wall)} in the scalar manifold, where in fact a
stable $2$-center BH configuration may decay into two single-center
constituents, whose scalar flows then separately evolve according to the
corresponding attractor dynamics.

The study of these phenomena has recently progressed in many
directions. By combining properties of $\mathcal{N}=2$ supergravity
and superstring theory, a number of interesting phenomena, such as
split flow tree, entropy enigma, bound state recombination walls,
and microstate counting have been investigated (see \textit{e.g.}
\cite{KSS-1}-\nocite
{DM-1,G-1,WC-1,Gimon-1,CS-1,David-1,Manschot-1,GP-1,WC-2,Sen}\cite
{Manschot-2}; as examples of earlier studies, see \textit{e.g.}
\cite{Rahmfeld}).

In the supergravity approximation, the detailed study of the split attractor
flow is made possible, in the limit of large (continuous) charges, by the
powerful restrictions imposed by electric-magnetic duality ($U$-duality%
\footnote{%
Here $U$-duality is referred to as the ``continuous'' limit (valid for large
values of the charges) of the non-perturbative string theory symmetries
introduced by Hull and Townsend in \cite{HT}.}).

An important ingredient in the study of attractor solutions in supergravity
is the concept of \textit{duality charge orbits}, and of the duality
invariants associated to them. In the past, a number of studies has led to a
rather complete classification of charge orbits for single-center extremal
BHs, and of their supersymmetry-preserving properties characterising the
corresponding BH background \cite{FG-1}-\nocite
{LPS-1,DFL-0-brane,BFGM1,ADFT-rev,Kallosh-rev,BNS-1,CFMZ1}\cite{BFGM2}.

After \cite{FGK}, it is known that in generic $\mathcal{N}=2$ theories all
scalar fields (belonging to Abelian vector multiplets) are stabilized in
terms of the charges in the near-horizon extremal BH geometry corresponding
to a ($\frac{1}{2}$-)BPS attractor configuration. On the other hand, for $%
\mathcal{N}=2$ non-BPS attractors, as well as for both BPS and non-BPS
attractors in $\mathcal{N}>2$-extended theories, not all scalar fields are
stabilized at the BH event horizon \cite{ADF-U-duality-d=4}, and \textit{%
``moduli spaces''} of attractor solutions exist \cite{Ferrara-Marrani-1-2}.

In $d=4$ supergravity theories, the fluxes of the two-form Abelian
field-strengths and their duals fit into the relevant (symplectic) irrepr. $%
\mathbf{R}$ of the $U$-duality group $G_{4}$. When considering a(n extremal)
$1$-center black $0$-brane (BH) background, such fluxes are referred to as
electric and magnetic black hole charges. The irrepr. charge space $\mathbf{R%
}$ exhibits a stratification in terms of disjoint orbits, each of them
supporting a distinct class of $1$-center BH solutions \cite{FG-1}-\nocite
{LPS-1,DFL-0-brane,BFGM1,ADFT-rev,Kallosh-rev,BNS-1,CFMZ1}\cite{BFGM2}.
Within theories with symmetric coset scalar manifolds $\frac{G_{4}}{H_{4}}$
(where $H_{4}$ is the maximal symmetric subgroup of $G_{4}$), a unique
duality-invariant polynomial of the charge irrepr. $\mathbf{R}$ of $G_{4}$
(in which the charges of a $1$-center BH sit) exists.

This is a \textit{quadratic} polynomial $\mathcal{I}_{2}$ in $\mathcal{N}=2$
symmetric special K\"{a}hler (SK) geometries with vanishing $C$-tensor (%
\textit{minimal coupling} sequence \cite{Luciani}). The same is also true
for the general $\mathcal{N}=3$ theory \cite{N=3} and for \textit{``pure''} $%
\mathcal{N}=4$ supergravity \cite{CSF} (the so-called \textit{axion-dilaton}
model, whose truncation down to $\mathcal{N}=2$ gives rise to the $\mathbb{CP%
}^{1}$ model - first element of the \textit{minimal coupling} sequence - in
a non-manifestly $U\left( 1,1\right) $-covariant symplectic basis).

On the other hand, symmetric $d$-special K\"{a}hler geometries (for a
comprehensive treatment, see \textit{e.g.} \cite{dWVVP}), based on degree-$3$
Euclidean Jordan algebras \cite{GST1,GST2}, have a unique duality-invariant
polynomial $\mathcal{I}_{4}$ which is \textit{quartic} in charges. Some of
these theories correspond to certain classical limits of moduli spaces of
Calabi-Yau internal manifolds in superstring compactifications \cite{David-1}%
. In particular, the simplest $d$-SK geometry, namely the symmetric $t^{3}$
model (see Sec. \ref{Gen-Analysis-t^3}, and Refs. therein), pertains to the
volume modulus in the large volume limit of compactifications of Type II
superstrings on Calabi-Yau threefolds.

For $p$-center (extremal) BHs, the study of charge orbits and duality
invariants is not known yet. The unique exception is given by the above
mentioned \textit{minimal coupling} $\mathbb{CP}^{n}$ sequence; indeed, the $%
2$-center dynamics, marginal stability and the properties of the related
split attractor flows have been recently shown to depend on four $U$-duality
invariants in \cite{MS-FMO-1} (in the same paper, a generalisation to p($%
\geqslant 3$)-center solutions was indicated, as well).\medskip

The present investigation is devoted to the study of duality charge orbits
supporting $2$-center extremal BHs in $\mathcal{N}=2$ symmetric $d$-SK
geometry based on the so-called Jordan symmetric infinite sequence \cite
{GST2}
\begin{equation}
\mathcal{N}=2:\frac{SL\left( 2,\mathbb{R}\right) }{U\left( 1\right) }\times
\frac{SO\left( 2,n\right) }{SO\left( 2\right) \times SO\left( n\right) }%
~\left( \mathbb{R}\oplus \mathbf{\Gamma }_{1,n-1}\right) ,
\label{N=2-J-symm-seq}
\end{equation}
where the round brackets in the right-hand side denote the corresponding
\textit{reducible} degree-$3$ Euclidean Jordan algebras \cite{GST1,GST2}
(see also \textit{e.g.} \cite{G-SAM-Lect}, and Refs. therein). In
particular, we will focus on the symmetric minimal rank-3 $stu$ model \cite
{Duff-stu,BKRSW} (which is a sub-sector of all symmetric $d$-SK geometries)
and its lower-rank descendants, namely the $st^{2}$ and $t^{3}$ models.

In general, in presence of a $p$ center (extremal) BH solution ($p\in
\mathbb{N}$), the number $I_{p}$ of independent $G_{4}$-invariant
polynomials built out with $p$ distinct copies of the charge irrepr. $%
\mathbf{R}$ of $G_{4}$ is given by the formula
\begin{equation}
p\text{dim}_{\mathbb{R}}\mathbf{R}=\text{dim}_{\mathbb{R}}\mathcal{O}%
_{p}+I_{p},  \label{general-1}
\end{equation}
where
\begin{equation}
\mathcal{O}_{p}\equiv \frac{G_{4}}{\mathcal{H}_{4,p}}  \label{charge-orbit-p}
\end{equation}
is the relevant $p$-center charge orbit, spanned by a vector of fluxes of
real dimension $p$dim$_{\mathbb{R}}\mathbf{R}$. In general, the counting of $%
I_{p}$ given by the formul\ae\ (\ref{general-1})-(\ref{charge-orbit-p})
depends only on the compact form of the symmetry groups involved, and thus
it is not affected by the supersymmetry properties exhibited by the
corresponding $p$-center BH background.

For example, in the case of BPS $p$-center extremal BHs in $\mathcal{N}=2$
\textit{minimally coupled} supergravity \cite{Luciani}, one obtains ($%
p\leqslant n+1$; see Sec. 4.2.1 of \cite{MS-FMO-1})
\begin{equation}
\left.
\begin{array}{r}
p\text{dim}_{\mathbb{R}}\mathbf{R}=2\left( n+1\right) p; \\
\\
\mathcal{O}_{BPS,p}=\frac{U\left( 1,n\right) }{U\left( n+1-p\right) }
\end{array}
\right\} \Rightarrow I_{p}=p^{2}.  \label{ressss}
\end{equation}

A new phenomenon occurring when $p>1$ is the fact that the various $G_{4}$%
-invariant polynomials arrange into irreprs. (multiplets) of an
``horizontal'' symmetry group, encoding the combinatoric structure of the $p$%
-center solutions of the theory. In the $\mathcal{N}=2$ \textit{minimally
coupled} theory, such an ``horizontal'' group is given by $U_{h}\left(
p\right) $ \cite{MS-FMO-1} (the subscript ``$h$'' stands for ``horizontal''
throughout). On the other hand, for the cubic models considered in the
present paper it is\footnote{%
Actually, in these cases the horizontal symmetry group is $GL\left( p,%
\mathbb{R}\right) $, where the additional scale symmetry with respect to $%
SL\left( p,\mathbb{R}\right) $ is encoded by the homogeneity of the $G_{4}$%
-invariant polynomials in charges.} $SL_{h}\left( p,\mathbb{R}\right) $ (see
Sec. \ref{Hor-Symm}).

For all $\mathcal{N}=2$ theories, dim$_{\mathbb{R}}\mathbf{R}=2n_{V}+2$,
where $n_{V}$ is the number of Abelian vector multiplets coupled to the
gravity multiplets. Thus, at least for $\mathcal{N}=2$ \textit{symmetric}
coset vector multiplets' scalar manifolds, the dimension of a ``large''
charge orbit $\mathcal{O}_{p=1}$ reads
\begin{equation}
\text{dim}_{\mathbb{R}}\mathcal{O}_{p=1}=2n_{V}+1=\text{dim}_{\mathbb{R}%
}\left( \frac{G_{4}}{H_{4,0}}\right) =\text{dim}_{\mathbb{R}}\mathbf{R}-1,
\label{p=1-large-N=2}
\end{equation}
where $H_{4}=H_{4,0}\times U\left( 1\right) $ is the maximal compact
subgroup of $G_{4}$, as well as the stabilizer of the scalar manifold
itself. Thus, the application of general relation (\ref{general-1}) to the $%
\mathcal{N}=2$ $1$-center case (\ref{p=1-large-N=2}) (which can be traced
back to the very structure of SK geometry \cite{FG-1,BFGM1}) yields to the
well known result
\begin{equation}
I_{p=1}=\text{dim}_{\mathbb{R}}\mathbf{R-}\text{dim}_{\mathbb{R}}\mathcal{O}%
_{p=1}=1,
\end{equation}
and the ``large'' nature of $\mathcal{O}_{p=1}$ means that it supports a
non-vanishing value of the unique $G_{4}$-invariant.

As mentioned above, we will focus on the $stu$, $st^{2}$ and $t^{3}$ $d$-SK
geometries, respectively corresponding to the rank-$3$, rank-$2$ and rank-$1$
symmetric cosets (see the treatment of Secs. \ref{Gen-Analysis-stu}, \ref
{Gen-Analysis-(1/2)st^2} and \ref{Gen-Analysis-t^3} for more detail, and
Refs. Therein):
\begin{equation}
\underset{\mathcal{F}=stu}{\left[ \frac{SL\left( 2,\mathbb{R}\right) }{%
U\left( 1\right) }\right] ^{3}},\text{~}\underset{\mathcal{F}=st^{2}}{\left[
\frac{SL\left( 2,\mathbb{R}\right) }{U\left( 1\right) }\right] ^{2}},\text{~}%
\underset{\mathcal{F}=t^{3}}{\frac{SL\left( 2,\mathbb{R}\right) }{U\left(
1\right) }}.
\end{equation}
The charge irrepr. $\mathbf{R}$ respectively is the $\left( \mathbf{2},%
\mathbf{2},\mathbf{2}\right) $ (spin $s=\left( \frac{1}{2},\frac{1}{2},\frac{%
1}{2}\right) $) of $\left[ SL\left( 2,\mathbb{R}\right) \right] ^{3}$, the $%
\left( \mathbf{3},\mathbf{2}\right) $ (spin $s=\left( 1,\frac{1}{2}\right) $%
) of $\left[ SL\left( 2,\mathbb{R}\right) \right] ^{2}$, and the $\left(
\mathbf{4}\right) $ (spin $s=\frac{3}{2}$) of $\left[ SL\left( 2,\mathbb{R}%
\right) \right] $ (see also the discussion in Sec. 5 of \cite{BMOS-1}). For
these models, the generic (``large'') $p=2$-center charge orbit $\mathcal{O}%
_{p=2}$ has no continuous stabilizer, so it just coincides with $G_{4}$
itself. Thus, the application of the general formula (\ref{general-1}) with $%
p=2$ in the theories under consideration yields that
\begin{equation}
I_{p=2}=\underset{stu^{{}}}{7},~\underset{st^{2}}{6},~\underset{t^{3}}{5}.
\label{counting-ind-invs}
\end{equation}

As discussed in Sec. \ref{T-tensor} within the (manifestly $G_{4}$%
-covariant) so-called Calabi-Vesentini\footnote{%
The Calabi-Vesentini basis for charges and holomorphic sections is
discussed in App. \ref{App-t^3}.} basis
\cite{CV-original,CDFVP-1,BKRSW}, a remarkable property of the $stu$
and $st^{2}$ models is that $G_{4}$ is \textit{reducible} (namely,
factorised: $G_{4}=\left[ SL\left( 2,\mathbb{R}\right) \right] ^{3}$ for $%
stu $, and $G_{4}=\left[ SL\left( 2,\mathbb{R}\right) \right] ^{2}$ for $%
st^{2}$). This generally allows for the existence of more independent $G_{4}$%
-invariant polynomials with respect to symmetric theories with irreducible $%
G_{4}$ (such as the $\mathcal{N}=2$ ``magic'' models \cite{GST1,GST2}).
Actually, both the whole $\mathcal{N}=2$ Jordan symmetric sequence (\ref
{N=2-J-symm-seq}) (whose the $stu$ and $st^{2}$ models are the $n=2$ and $%
n=1 $ element, respectively) and the sequence pertaining to
$\mathcal{N}=4$ supergravity (see Eq. (\ref{N=4}) further below)
have \textit{factorised} scalar manifolds, and the aforementioned
property (as well as the possibility to perform a Calabi-Vesentini
manifestly $G_{4}$-covariant treatment) extends to these two
infinite sequences. It is here worth pointing out that, for
symmetric $d$-SK geometries, the reducible (irreducible) nature of
$G_{4}$ is ultimately due to the reducibility (irreducibility) of
the underlying rank-$3$ Euclidean Jordan algebra (for the reducible
cases, see Eqs. (\ref{N=2-J-symm-seq}) and (\ref{N=4})).

For the $stu$ model, in Sec. \ref{Gen-Analysis-stu} we will show that there
is a basis of seven independent $\left[ SL\left( 2,\mathbb{R}\right) \right]
^{3}$-invariant homogeneous polynomials, six of them are quartic and the one
is quadratic in charges. Within the notation specified in Sec. \ref{Hor-Symm}%
, these polynomial invariants arrange as follows:
\begin{equation}
\begin{array}{l}
\mathcal{W}\equiv \left\langle \mathcal{Q}_{1},\mathcal{Q}_{2}\right\rangle
\mathcal{~}\text{\textit{quadratic}~in~charges}~\text{(see Eq. (\ref
{SymplProd-stu}))}; \\
\\
\left( \frak{I},~\mathcal{X}\right) ~\text{\textit{quartic}~in~charges}~%
\text{(see Eq. (\ref{InvVec}))},
\end{array}
\label{1+5+1}
\end{equation}
where $\frak{I}$ is a quintet of five invariants, and $\left\langle \mathcal{%
Q}_{1},\mathcal{Q}_{2}\right\rangle \mathcal{\ }$denotes the symplectic
product of the charge vectors pertaining to the two centers.

For the $st^{2}$ and $t^{3}$ models, in which the number of independent $%
G_{4}$-invariants is smaller, we will exhibit polynomial constraints,
manifestly invariant under the aforementioned ``horizontal'' symmetry $%
SL_{h}\left( 2,\mathbb{R}\right) $, which relate such invariants. We
anticipate that both $\mathcal{W}$ and $\mathcal{X}$ of Eq. (\ref{1+5+1})
are singlets under $SL_{h}\left( 2,\mathbb{R}\right) $, whereas $\frak{I}$
sit in an irrepr. $\mathbf{5}$ (spin $s=2$) of $SL_{h}\left( 2,\mathbb{R}%
\right) $ itself (see(\ref{InvVec})). In the $stu$ model, the relevant
polynomial constraint has the (order-$12$ in charges) structure (see Eq. (%
\ref{P12=0-2}) for explicit form):
\begin{equation}
\mathcal{P}_{12}\left( \mathbf{I}_{6},\mathcal{W},\mathcal{X},\text{Tr}%
\left( \frak{I}^{2}\right) ,\text{Tr}\left( \frak{I}^{3}\right) \right) =0,
\label{PP12=0}
\end{equation}
which always allows one \textit{e.g.} to eliminate Tr$\left( \frak{I}%
^{3}\right) $ in terms of $\mathbf{I}_{6}$, and \textit{vice versa}. As
detailed in App. \ref{App-stu}, the further reduction to the $st^{2}$ model
gives rise to a polynomial of order-$16$ in charges structure (see Eqs. (\ref
{P16=0-2}) and (\ref{P16-t^3}) for explicit form):
\begin{equation}
\mathcal{P}_{16}\left( \mathcal{W},\mathcal{X},\text{Tr}\left( \frak{I}%
^{2}\right) ,\text{Tr}\left( \frak{I}^{3}\right) \right) =0,  \label{PP16=0}
\end{equation}
and it can be regarded a fourth order algebraic equation for $\mathcal{W}%
^{2} $. In the $t^{3}$ model, a relation of type (\ref{PP16=0}) (with $%
\mathcal{P}_{16}$ given by (\ref{P16-t^3})) also holds, together with the
further constraint $\mathcal{X}=0$ (see Eq. (\ref{t^3-rell})), of order $4$
in charges. Note that the quintet $\frak{I}$ enters Eqs. (\ref{PP12=0}) and (%
\ref{PP16=0}) only through the $SL_{h}\left( 2,\mathbb{R}\right) $-invariant
expressions Tr$\left( \frak{I}^{2}\right) $ and Tr$\left( \frak{I}%
^{3}\right) $, respectively given by Eqs. (\ref{Tr(I^2)}) and (\ref{Tr(I^3)}%
).

Alternatively, by using (\ref{PP12=0}), one can replace Tr$\left( \frak{I}%
^{3}\right) $ with $\mathbf{I}_{6}$ (defined in (\ref{sextic})) as generator
of a complete lowest-degree basis (\ref{4-2}) of manifestly $SL_{h}\left( 2,%
\mathbb{R}\right) $-invariant polynomials. Thus, Eq. (\ref{PP16=0}) gets
replaced by the degree-$8$ constraint (\ref{P8=0-2}), which is nothing but
the vanishing of the determinant of a suitably defined Gramian matrix $%
\mathbf{G}$ (defined by (\ref{G-matrix})-(\ref{G-matrix-2})).\bigskip\

The paper is organised as follows.

In Sec. \ref{Gen-Analysis-stu} we recall the $stu$ model and its properties
(in the ``special coordinates'' symplectic frame).

In Sec. \ref{T-tensor} we introduce a general formalism for the
construction and analysis of polynomial $G_{4}$-invariants in all
cases in which $G_{4}$ is factorised. This formalism is worked out
in the Calabi-Vesentini basis \cite{CDFVP-1}, and it is based on the
so-called $\mathbb{T}$-tensor; we also briefly outline the relation
between the $\mathbb{T}$-tensor and the corresponding counterpart
for irreducible cubic geometries. In particular, this formalism
applies to $stu$ and $st^{2}$ since they are, as mentioned above,
the $n=2$ and $n=1$ element of the $\mathcal{N}=2$ Jordan symmetric
sequence (\ref{N=2-J-symm-seq}), respectively. The application to the rank-$%
1 $ \textit{irreducible} $t^{3}$ model deserves a separate treatment, given
in App. \ref{App-t^3}.

Sec. \ref{Hor-Symm} analyzes the crucial role played by the ``horizontal''
symmetry $SL_{h}\left( 2,\mathbb{R}\right) $ (generalisable to $SL_{h}\left(
p,\mathbb{R}\right) $ for $p\geqslant 3$ centers) in classifying the
polynomial $G_{4}$-invariants and in determining the structure of the
polynomial constraints relating them. In particular, for each order of
homogeneity in charges, the various $G_{4}$-invariants arrange into irreprs.
(multiplet) of the ``horizontal'' symmetry itself.

In Sec. \ref{stu-Descendants} the issue of independence (primitivity) of the
$G_{4}$-invariant in the models under consideration is addressed. Besides
the explicit computation based on the analysis of the rank of a suitably
defined Jacobian matrix, also the general counting argument based on formula
(\ref{general-1}) is given. A polynomial constraint of degree $12$ in
charges, involving also the unique $G_{4}$-invariant polynomial of order six
in charges (singlet under $SL_{h}\left( 2,\mathbb{R}\right) $) is given (and
derived in detail in App. \ref{App-stu}).

Then, in Secs. \ref{Gen-Analysis-(1/2)st^2} and \ref{Gen-Analysis-t^3} the
reduction of the $stu$ to $st^{2}$ respectively $t^{3}$ model is performed,
and in App. \ref{App-stu} the corresponding hierarchy of manifestly $%
SL_{h}\left( 2,\mathbb{R}\right) $-invariant polynomial constraints
(consistent with the result (\ref{counting-ind-invs})) is derived.

In Sec. \ref{SO_h(2,2)} we develop further the analysis of invariant
polynomials, by combining the ``horizontal'' symmetry $SL_{h}\left( 2,%
\mathbb{R}\right) $ with the ``vertical'' symmetry $SL_{v}\left( 2,\mathbb{R}%
\right) $. This latter, for the models treated in the present paper, is part
of the $d=4$ $U$-duality group $G_{4}$. Then, we use the characteristic
equation of the Gramian matrix $\mathbf{G}$ to exploit a manifestly $\left[
SL_{h}\left( 2,\mathbb{R}\right) \times SL\left( 2,\mathbb{R}\right) \right]
$-invariant formalism, actually holding for both the infinite \textit{%
reducible} sequences (\ref{N=2-J-symm-seq}) and (\ref{N=4}) of symmetric
scalar manifolds.

Finally, in Sec. \ref{Gen-Reducible-N=2-N=4} the extension of the previous
analysis to generic elements of the $\mathcal{N}=2$ Jordan symmetric and $%
\mathcal{N}=4$ \textit{reducible} infinite sequences is discussed; for the $%
\mathcal{N}=2$ sequence with $n\geqslant 3$ and for the whole $\mathcal{N}=4$
sequence ($n\geqslant 0$), the treatment is analogous, and the results
identical, to the case of the $stu$ model considered in Sec. \ref
{Gen-Analysis-stu}.

Three Appendices conclude the paper. In App. \ref{App-stu} we give
details on the derivation of the relevant polynomial constraints in
$stu$, $st^{2}$ and $t^{3}$ models. App. \ref{App-t^3} discusses the
relation between the usual ``special coordinates'' symplectic basis
(used in $D$-brane description) and the Calabi-Vesentini basis. App.
\ref{Proof-Completeness} presents a complete basis for the $SO\left( n,\mathbb{C%
}\right) $-invariant polynomials, a rigorous result mentioned in
Sec. \ref{SO_h(2,2)}.
\bigskip\

We should point out that, although we perform an analysis for BPS
(``large'') multi-center extremal BHs, the extension to non-BPS ``large'' as
well as to ``small'' BHs is straightforward. Strictly speaking, it is worth
recalling that, at the best of our current understanding (see \textit{e.g.}
\cite{MS-FM-1}), the marginal decay and split attractor flow can be
generalised to $\mathcal{N}=2$ non-BPS cases only with $\mathcal{I}_{4}>0$
(namely, the $Z_{H}=0$ attractors). Anyhow, the analysis of $p$-center
charge orbits can be carried out for all cases (see the comments below Eq. (%
\ref{charge-orbit-p})).

\section{\label{Gen-Analysis-stu}The $stu$ Model}

We start and consider the so-called $\mathcal{N}=2$, $d=4$ $stu$ model \cite
{Duff-stu,BKRSW}. In the ``special coordinates'' basis (see \textit{e.g.}
\cite{CDF-rev} and Refs. therein), this model is defined by the prepotential
\begin{equation}
F\left( X\right) \equiv \frac{X^{1}X^{2}X^{3}}{X^{0}}=\frac{1}{3!}d_{ijk}%
\frac{X^{i}X^{j}X^{k}}{X^{0}}\Leftrightarrow d_{123}=1,  \label{F-def-stu}
\end{equation}
thus implying
\begin{eqnarray}
F_{0} &=&\frac{\partial F}{\partial X^{0}}=-X^{0}\mathcal{F};~F_{1}=\frac{%
\partial F}{\partial X^{1}}=X^{0}\mathcal{F}_{1};~F_{2}=\frac{\partial F}{%
\partial X^{2}}=X^{0}\mathcal{F}_{2};~F_{3}=\frac{\partial F}{\partial X^{3}}%
=X^{0}\mathcal{F}_{3}; \\
\mathcal{F} &\equiv &stu;~\mathcal{F}_{1}\equiv tu=\frac{\partial \mathcal{F}%
}{\partial s};~\mathcal{F}_{2}\equiv su=\frac{\partial \mathcal{F}}{\partial
t};~\mathcal{F}_{3}\equiv st=\frac{\partial \mathcal{F}}{\partial u},
\end{eqnarray}
where
\begin{equation}
s\equiv \frac{X^{1}}{X^{0}},~t\equiv \frac{X^{2}}{X^{0}},~u\equiv \frac{X^{3}%
}{X^{0}}  \label{s-t-u-def}
\end{equation}
are the \textit{projective} coordinates. Through the definition (\ref
{s-t-u-def}), the $Sp\left( 8,\mathbb{R}\right) $-vector of holomorphic
symplectic sections can thus be written as follows:
\begin{equation}
\mathbf{V}\equiv \left(
\begin{array}{c}
X^{0} \\
X^{1} \\
X^{2} \\
X^{3} \\
F_{0} \\
F_{1} \\
F_{2} \\
F_{3}
\end{array}
\right) =\left(
\begin{array}{c}
1 \\
s \\
t \\
u \\
-\frac{1}{2}st^{2} \\
tu \\
su \\
st
\end{array}
\right) X^{0}=\left(
\begin{array}{c}
1 \\
s \\
t \\
u \\
-\mathcal{F} \\
\mathcal{F}_{1} \\
\mathcal{F}_{2} \\
\mathcal{F}_{3}
\end{array}
\right) X^{0}.  \label{s-t-u-sympl-sects}
\end{equation}
Here we will not report a detailed treatment of the $stu$ model (we address
the reader \textit{e.g.} to \cite{Duff-stu,BKRSW,BMOS-1,stu-unveiled,CDFY-2}%
), we will just confine ourselves to some basics, useful for the
developments given below.

The $stu$ model is based on the rank-$3$ completely factorised symmetric
coset
\begin{equation}
\frac{G_{4}}{H_{4}}=\frac{SL\left( 2,\mathbb{R}\right) }{U(1)}\times \frac{%
SO\left( 2,2\right) }{SO\left( 2\right) \times SO\left( 2\right) }\sim \left[
\frac{SL\left( 2,\mathbb{R}\right) }{U(1)}\right] ^{3},  \label{stu-stu}
\end{equation}
where $G_{4}=\left[ SL\left( 2,\mathbb{R}\right) \right] ^{3}$ is the $d=4$ $%
U$-duality group, and $H_{4}=\left[ U\left( 1\right) \right] ^{3}$ its
maximal compact subgroup (\textit{mcs}). This coset is the second element ($%
n=2$) of the aforementioned $\mathcal{N}=2$, $d=4$ Jordan symmetric sequence
(see \textit{e.g.} \cite{CFG,dWVVP}, and Refs. therein).

This model admits all classes of extremal BH attractors \cite{AM-Refs,FGK},
namely $\frac{1}{2}$-BPS, non-BPS $Z_{H}\neq 0$ and non-BPS $Z_{H}=0$ ones.
The BPS solutions were known after \cite{BKRSW,Shmakova}, whereas the
explicit expression of the non-BPS $Z_{H}=0$ attractors have been obtained
in \cite{BMOS-1}. The non-BPS $Z_{H}\neq 0$ attractor solutions were
obtained in full generality in \cite{stu-unveiled} (see also Refs. therein,
as well as \cite{CDFY-2}).

By introducing the $Sp\left( 8,\mathbb{R}\right) $-vector of magnetic and
electric charges (the naught index pertains to the graviphoton throughout)
\begin{equation}
\mathcal{Q}\equiv \left(
p^{0},p^{1},p^{2},p^{3},q_{0},q_{1},q_{2},q_{3}\right) ^{T},
\end{equation}
in the ``special coordinate basis'' the unique polynomial invariant
(homogeneous and quartic in the charges) of the $\left( \mathbf{2,2,2}%
\right) $ (namely spin $s=\left( \frac{1}{2},\frac{1}{2},\frac{1}{2}\right) $%
) irrepr. of $G_{4}=\left[ SL\left( 2,\mathbb{R}\right) \right] ^{3}$ reads
%
\begin{eqnarray}
\mathcal{I}_{4}\left( \mathcal{Q}\right) &\equiv &-\left( p^{0}\right)
^{2}q_{0}^{2}-\left( p^{1}\right) ^{2}q_{1}^{2}-\left( p^{2}\right)
^{2}q_{2}^{2}-\left( p^{3}\right) ^{2}q_{3}^{2}  \notag \\
&&-2p^{0}q_{0}p^{1}q_{1}-2p^{0}q_{0}p^{2}q_{2}-2p^{0}q_{0}p^{3}q_{3}+2p^{1}q_{1}p^{2}q_{2}+2p^{1}q_{1}p^{3}q_{3}+2p^{2}q_{2}p^{3}q_{3}
\notag \\
&&+4q_{0}p^{1}p^{2}p^{3}-4p^{0}q_{1}q_{2}q_{3}=-\text{Det}\left( \psi
\right) ,  \label{I4-stu}
\end{eqnarray}
where Det$\left( \psi \right) $ is the so-called Cayley's hyperdeterminant
\cite{Cayley}. $\mathcal{I}_{4}>0$ for $\frac{1}{2}$-BPS and non-BPS $%
Z_{H}=0 $, while $\mathcal{I}_{4}<0$ for non-BPS $Z_{H}\neq 0$ attractor
solutions, respectively (see Appendix II of \cite{BFGM1}). (\ref{I4-stu})
can be obtained from the general formula (for symmetric $d$-SK geometries;
see \cite{Ferrara-Gimon} for notation and further elucidation)
\begin{equation}
\mathcal{I}_{4}\left( \mathcal{Q}\right) =-\left(
p^{0}q_{0}+p^{i}q_{i}\right) ^{2}+\frac{2}{3}q_{0}d_{ijk}p^{i}p^{j}p^{k}-%
\frac{2}{3}p^{0}d^{ijk}q_{i}q_{j}q_{k}+d_{ijk}d^{ilm}p^{j}p^{k}q_{l}q_{m}\,,
\label{I4}
\end{equation}
by specifying $d_{123}=1=d^{123}$, consistently with the non-linear relation
(for symmetric $d$-SK geometries \cite{GST2,CVP})
\begin{equation}
d_{r(pq}d_{ij)k}d^{rkl}=\frac{4}{3}\delta _{(p}^{l}d_{qij)}.
\label{symmetric-cond}
\end{equation}
At the level of $1$-center quartic $G_{4}$-invariant polynomials, the
progressive reduction \textit{``}$stu\rightarrow st^{2}\rightarrow t^{3}$%
\textit{''} procedure has been discussed in Sect. 5 of \cite{BMOS-1}.

\section{\label{T-tensor}$2$-Center $G_{4}$-Invariants and The $\mathbb{T}$%
-tensor Formalism}

Let us now consider a double-center extremal BH in the $stu$ model, with the
charge vectors associated to the two centers respectively reading
\begin{eqnarray}
\mathcal{Q}_{1} &\equiv &\left(
p^{0},p^{1},p^{2},p^{3},q_{0},q_{1},q_{2},q_{3}\right) ^{T};  \label{Q1-stu}
\\
\mathcal{Q}_{2} &\equiv &\left(
P^{0},P^{1},P^{2},P^{3},Q_{0},Q_{1},Q_{2},Q_{3}\right) ^{T}.  \label{Q2-stu}
\end{eqnarray}
By switching to the so-called Calabi-Vesentini basis
\cite{CDFVP-1,BKRSW},
in the $stu$ model (and, as we will see below, in the related $st^{2}$ and $%
t^{3}$ model, as well) the analysis of the multi-center $U$-invariant
polynomials can efficiently be performed by using the following quantity,
which we dub ``$\mathbb{T}$-tensor'' ($\Lambda =0,1,2,3$):
\begin{eqnarray}
\mathbb{T}_{12} &\equiv &T_{\Lambda \Sigma }\left( \mathcal{Q}_{1}\mathcal{Q}%
_{2}\right) \equiv \frac{1}{2}\left( p_{\Lambda }Q_{\Sigma }-q_{\Lambda
}P_{\Sigma }+P_{\Lambda }q_{\Sigma }-Q_{\Lambda }p_{\Sigma }\right) ;
\label{T12-pre} \\
T_{\Lambda \Sigma }\left( \mathcal{Q}_{1}\mathcal{Q}_{2}\right)
&=&-T_{\Sigma \Lambda }\left( \mathcal{Q}_{1}\mathcal{Q}_{2}\right)
=T_{\Lambda \Sigma }\left( \mathcal{Q}_{2}\mathcal{Q}_{1}\right) =-T_{\Sigma
\Lambda }\left( \mathcal{Q}_{2}\mathcal{Q}_{1}\right) ,
\end{eqnarray}
where we understand the raising and lowering of $\Lambda $-indices to be
done with the metrics $\eta _{\Lambda \Sigma }$ and $\eta ^{\Lambda \Sigma }$
of $SO\left( 2,2\right) $. Note that the $1$-center limit $1\equiv 2$ of (%
\ref{T12-pre}) consistently yields the antisymmetric rank-$2$ tensors
usually considered in the $1$-center analysis (see \textit{e.g.} \cite
{Ferrara-Maldacena,ADF-U-duality-d=4,CFMZ1,ADFT-FO-1})
\begin{eqnarray}
\mathbb{T}_{1} &\equiv &T_{\Lambda \Sigma }\left( \mathcal{Q}_{1}^{2}\right)
\equiv p_{\Lambda }q_{\Sigma }-q_{\Lambda }p_{\Sigma };  \label{T1-pre} \\
\mathbb{T}_{2} &\equiv &T_{\Lambda \Sigma }\left( \mathcal{Q}_{2}^{2}\right)
\equiv P_{\Lambda }Q_{\Sigma }-Q_{\Lambda }P_{\Sigma }.  \label{T2-pre}
\end{eqnarray}

While the charges in the ``special coordinates'' basis (namely, the ones
used in Eq. (\ref{I4-stu})) are manifestly covariant only with respect to
the $d=5$ $U$-duality group $G_{5}=\left[ SO\left( 1,1\right) \right] ^{2}$,
the Calabi-Vesentini basis is manifestly covariant under the whole $d=4$ $U$%
-duality group $SL\left( 2,\mathbb{R}\right) \times SO\left( 2,2\right) $
\cite{CDFVP-1}. In the latter basis, by virtue of the factorised nature of
the $U$-duality group, the charge vector $\mathcal{Q}$ splits into a
magnetic-electric $SL\left( 2,\mathbb{R}\right) $-doublet of $SO\left(
2,2\right) $ vectors, as follows:
\begin{equation}
\mathcal{Q}_{\mathbb{A}}=\left( p^{\Lambda },q_{\Lambda }\right) \equiv
\left( Q_{1\Lambda },Q_{2\Lambda }\right) \equiv \mathcal{Q}_{\alpha \Lambda
},  \label{Q-split}
\end{equation}
where $\alpha =1,2$ is in the fundamental $\mathbf{2}$ (spin $s=1/2$)
irrepr. of $SL\left( 2,\mathbb{R}\right) $, and $\Lambda $ is in the $%
\mathbf{4}$ vector irrepr. of $SO\left( 2,2\right) $. As a consequence, by
defining
\begin{equation}
p^{2}\equiv p^{\Lambda }p^{\Sigma }\eta _{\Lambda \Sigma },~q^{2}\equiv
q_{\Lambda }q_{\Sigma }\eta ^{\Lambda \Sigma },~p\cdot q\equiv p^{\Lambda
}q_{\Lambda },
\end{equation}
the unique quartic $1$-center $G_{4}$-invariant polynomial
(\ref{I4-stu}) \cite{CY,Duff-stu,CT} can be rewritten as
follows\footnote{For reasons of covariance, in Eqs. (\ref{I4-CV}),
(\ref{I+2-CV})-(\ref{sextic}), (\ref{T-bold}), (\ref{W-call}),
(\ref{x-Frak}) and (\ref{t^3-rell}), ``Tr'' denotes the
$\eta$-trace, namely the trace in which the indices are raised and
lowered by the pseudo-Euclidean metric $\eta$.}
\begin{equation}
\mathcal{I}_{4}\left( \mathcal{Q}\right) \equiv p^{2}q^{2}-\left( p\cdot
q\right) ^{2}=\frac{1}{2}T_{\Lambda \Sigma }\left( \mathcal{Q}^{2}\right)
T_{\Xi \Omega }\left( \mathcal{Q}^{2}\right) \eta ^{\Lambda \Xi }\eta
^{\Sigma \Omega }=-\frac{1}{2}\text{Tr}\left( \mathbb{T}^{2}\right) .
\label{I4-CV}
\end{equation}

Due to the \textit{reducible} (factorised) nature of the $d=4$ $U$-duality
group $G_{4}$ in $stu$ and $st^{2}$ models, the $\mathbb{T}$-tensors $%
\mathbb{T}_{12}$, $\mathbb{T}_{1}$ and $\mathbb{T}_{2}$ (defined by (\ref
{T12-pre})-(\ref{T2-pre})) are the basic structures needed to analyse the $%
p\geqslant 2$-center $G_{4}$-invariant polynomials. Here below we give the
complete analysis of all non-vanishing (\textit{a priori}) independent
invariant polynomials constructed with all possible contractions of two and
three $\mathbb{T}$-tensors out of the ones defined by (\ref{T12-pre})-(\ref
{T2-pre}):

\begin{itemize}
\item  \textbf{two }$\mathbb{T}$\textbf{'s.} For $p=2$ centers, there are
six non-vanishing (\textit{a priori}) independent invariant polynomials
constructed with all possible contractions of two $\mathbb{T}$-tensors out
of the ones defined by (\ref{T12-pre})-(\ref{T2-pre}), namely (recall
definition (\ref{I4-CV})):
\begin{eqnarray}
\mathbf{I}_{+2}\left( \mathcal{Q}_{1}^{4}\right) &\equiv &\mathcal{I}%
_{4}\left( \mathcal{Q}_{1}\right) =-\frac{1}{2}\text{Tr}\left( \mathbb{T}%
_{1}^{2}\right) =p^{2}q^{2}-\left( p\cdot q\right) ^{2};  \label{I+2-CV} \\
\mathbf{I}_{+1}\left( \mathcal{Q}_{1}^{3}\mathcal{Q}_{2}\right) &\equiv &-%
\frac{1}{2}\text{Tr}\left( \mathbb{T}_{1}\mathbb{T}_{12}\right) =\frac{1}{2}%
\left[ p^{2}\left( q\cdot Q\right) +q^{2}\left( p\cdot P\right) -\left(
p\cdot q\right) \left( p\cdot Q\right) -\left( p\cdot q\right) \left( P\cdot
q\right) \right] ;  \label{I+1-CV} \\
\mathbf{I}^{\prime }\left( \mathcal{Q}_{1}^{2}\mathcal{Q}_{2}^{2}\right) &=&-%
\frac{1}{2}\text{Tr}\left( \mathbb{T}_{1}\mathbb{T}_{2}\right) =\left(
p\cdot P\right) \left( q\cdot Q\right) -\left( p\cdot Q\right) \left( q\cdot
P\right) ;  \label{I'-CV} \\
\mathbf{I}^{\prime \prime }\left( \mathcal{Q}_{1}^{2}\mathcal{Q}%
_{2}^{2}\right) &=&-\frac{1}{2}\text{Tr}\left( \mathbb{T}_{12}^{2}\right) =%
\frac{1}{4}\left[
\begin{array}{l}
p^{2}Q^{2}+q^{2}P^{2}+2\left( p\cdot P\right) \left( q\cdot Q\right) \\
-\left( p\cdot Q\right) ^{2}-\left( q\cdot P\right) ^{2}-2\left( p\cdot
q\right) \left( P\cdot Q\right)
\end{array}
\right] ;  \label{I''-CV} \\
\mathbf{I}_{-1}\left( \mathcal{Q}_{1}\mathcal{Q}_{2}^{3}\right) &\equiv &-%
\frac{1}{2}\text{Tr}\left( \mathbb{T}_{2}\mathbb{T}_{12}\right) =\frac{1}{2}%
\left[ P^{2}\left( q\cdot Q\right) +Q^{2}\left( p\cdot P\right) -\left(
P\cdot Q\right) \left( P\cdot q\right) -\left( P\cdot Q\right) \left( p\cdot
Q\right) \right] ;  \notag \\
&&  \label{I-1-CV} \\
\mathbf{I}_{-2}\left( \mathcal{Q}_{2}^{4}\right) &\equiv &\mathcal{I}%
_{4}\left( \mathcal{Q}_{2}\right) =-\frac{1}{2}\text{Tr}\left( \mathbb{T}%
_{2}^{2}\right) =P^{2}Q^{2}-\left( P\cdot Q\right) ^{2}.  \label{I-2-CV}
\end{eqnarray}

\item  \textbf{three }$\mathbb{T}$\textbf{'s.} For $p=2$ centers, there is
only one possible non-vanishing invariant polynomial constructed with all
possible contractions of three $\mathbb{T}$-tensors out of the ones defined
by (\ref{T12-pre})-(\ref{T2-pre}), namely:
\begin{equation}
\mathbf{I}_{6}\left( \mathcal{Q}_{1}^{3}\mathcal{Q}_{2}^{3}\right) \equiv -%
\text{Tr}\left( \mathbb{T}_{1}\mathbb{T}_{2}\mathbb{T}_{12}\right) =-\frac{1%
}{2}\left[
\begin{array}{l}
\left( q\cdot P\right) \left( p\cdot Q\right) ^{2}-\left( p\cdot Q\right)
\left( P\cdot q\right) ^{2} \\
+\left( q\cdot P\right) \left( q\cdot Q\right) \left( p\cdot P\right)
-\left( Q\cdot p\right) \left( q\cdot Q\right) \left( p\cdot P\right) \\
+\left( q\cdot P\right) \left( p\cdot q\right) \left( P\cdot Q\right)
-\left( Q\cdot p\right) \left( p\cdot q\right) \left( P\cdot Q\right) \\
-\left( q\cdot P\right) p^{2}Q^{2}+\left( Q\cdot p\right) P^{2}q^{2} \\
-\left( q\cdot Q\right) P^{2}\left( p\cdot q\right) +\left( q\cdot Q\right)
p^{2}\left( P\cdot Q\right) \\
-\left( p\cdot P\right) q^{2}\left( P\cdot Q\right) +\left( p\cdot P\right)
Q^{2}\left( p\cdot q\right) ,
\end{array}
\right] .  \label{sextic}
\end{equation}
This $G_{4}$-invariant polynomial will turn out to be dependent on the
lower-degrees $G_{4}$-invariant polynomials in all $\mathcal{N}=2$, $d=4$
models ($stu$, $st^{2}$ and $t^{3}$) which we consider in the present
investigation.
\end{itemize}

\section{\label{Hor-Symm}The Role of the Horizontal Symmetry $SL_{h}\left( 2,%
\mathbb{R}\right) $}

The rank-$2$ antisymmetric $\mathbb{T}$-tensors (\ref{T12-pre})-(\ref{T2-pre}%
)\ fit into an irrepr. $\mathbf{3}$ (spin $s=1$) of a further ``horizontal''
symmetry $SL_{h}\left( 2,\mathbb{R}\right) $, which takes into account the
combinatorics under the exchange of the centers $1\leftrightarrow 2$ (here
the subscript ``$h$'' stands for ``horizontal''). Such a $\mathbf{3}$
irrepr. is the symmetric part of the tensor product of two fundamental
irrepr. $\mathbf{2}$ (spin $s=1/2$) of $SL_{h}\left( 2,\mathbb{R}\right) $,
in which $\mathcal{Q}_{1}$ and $\mathcal{Q}_{2}$ sit, with helicity $+1/2$
and $-1/2$, respectively:
\begin{equation}
SL_{h}\left( 2,\mathbb{R}\right) :\mathbf{2}\times \mathbf{2}=\overset{%
\left( \mathbb{T}_{12},\mathbb{T}_{1},\mathbb{T}_{2}\right) }{\mathbf{3}_{s}}%
+\overset{\mathbb{T}_{a}}{\mathbf{1}_{a}},  \label{decomp-1}
\end{equation}
where
\begin{equation}
\mathbb{T}_{a,\Lambda \Sigma }\left( \mathcal{Q}_{1}\mathcal{Q}_{2}\right)
\equiv \frac{1}{2}\left(q_{\Lambda}P_{\Sigma }+q_{\Sigma }P_{\Lambda}-
p_{\Lambda }Q_{\Sigma }-p_{\Sigma }Q_{\Lambda }\right)  \label{Ta}
\end{equation}
is a rank-$2$ symmetric tensor, which is antisymmetric under $%
1\leftrightarrow 2$, and thus it vanishes for $1\equiv 2$. Note that under $%
1\leftrightarrow 2$ $\mathbb{T}_{12}$ is invariant, whereas $\mathbb{T}%
_{1}\leftrightarrow \mathbb{T}_{2}$.

From the definitions (\ref{I'-CV}) and (\ref{I''-CV}), the squared norm of
the $3$-vector $\mathbb{T}\equiv \left( \mathbb{T}_{1},\mathbb{T}_{12},%
\mathbb{T}_{2}\right) $ reads
\begin{equation}
\left\| \mathbb{T}\right\| ^{2}\equiv -\frac{1}{2}\text{Tr}\left( \mathbb{T}%
_{1}\mathbb{T}_{2}\right) +\frac{1}{2}\text{Tr}\left( \mathbb{T}%
_{12}^{2}\right) =\mathbf{I}^{\prime }-\mathbf{I}^{\prime \prime }.
\label{T-bold}
\end{equation}
This is a singlet of $SL_{h}\left( 2,\mathbb{R}\right) $, symmetric under
the center exchange $1\leftrightarrow 2$.

Also the subscripts of the the four $G_{4}$-invariants $\mathbf{I}_{+2}$, $%
\mathbf{I}_{+1}$, $\mathbf{I}_{-1}$ and $\mathbf{I}_{-2}$, defined by (\ref
{I+2-CV}), (\ref{I+1-CV}), (\ref{I-1-CV}) and (\ref{I-2-CV}), denote their
helicity with respect to the relevant irrepr. of the horizontal symmetry $%
SL_{h}\left( 2,\mathbb{R}\right) $. Indeed, by further defining
\begin{equation}
\mathbf{I}_{0}\equiv \frac{1}{3}\left( \mathbf{I}^{\prime }+2\mathbf{I}%
^{\prime \prime }\right) ,  \label{I0}
\end{equation}
the five $G_{4}$-invariants $\mathbf{I}_{+2}$, $\mathbf{I}_{+1}$, $\mathbf{I}%
_{0}$, $\mathbf{I}_{-1}$ and $\mathbf{I}_{-2}$ sit in the $\mathbf{5}$ (spin
$s=2$) irrepr. of $SL_{h}\left( 2,\mathbb{R}\right) $ itself:
\begin{equation}
\frak{I}\equiv \overset{\text{spin~}s=2}{\mathbf{5}}\equiv \left(
\begin{array}{ccccc}
\mathbf{I}_{+2}, & \mathbf{I}_{+1}, & \mathbf{I}_{0}, & \mathbf{I}_{-1}, &
\mathbf{I}_{-2}
\end{array}
\right) ;~Tr\frak{I}=0.  \label{InvVec}
\end{equation}
The very definitions (\ref{I+2-CV})-(\ref{I-2-CV}) and (\ref{I0})
characterize the $\mathbf{5}$ given in (\ref{InvVec}) as a part symmetric
tensor product of two irreprs. $\mathbf{3}$ of $SL_{h}\left( 2,\mathbb{R}%
\right) $ itself (in which the $\mathbb{T}$-tensors (\ref{T12-pre})-(\ref
{T2-pre}) sit):
\begin{equation}
SL_{h}\left( 2,\mathbb{R}\right) :\mathbf{3}\times \mathbf{3}=\overset{%
\left( \mathbf{I}_{+2},\mathbf{I}_{+1},\mathbf{I}_{0},\mathbf{I}_{-1},%
\mathbf{I}_{-2}\right) }{\mathbf{5}_{s}}+\overset{\mathbf{T}}{\mathbf{1}_{s}}%
+\mathbf{3}_{a}.  \label{3x3}
\end{equation}
Note that the $SL_{h}\left( 2,\mathbb{R}\right) $-singlet $\mathbf{T}$
defined in (\ref{T-bold}) sits in the $\mathbf{1}_{s}$ in the right-hand
side of decomposition (\ref{3x3}).

Notice that all the $G_{4}$-quartic invariants $\mathbf{I}_{+2}$, $\mathbf{I}%
_{+1}$, $\mathbf{I}_{0}$, $\mathbf{I}^{\prime }$, $\mathbf{I}^{\prime \prime
}$, $\mathbf{I}_{-1}$ and $\mathbf{I}_{-2}$ consistently reduce to $\mathcal{%
I}_{4}\left( \mathcal{Q}\right) $ defined in (\ref{I4-CV}) in the $1$-center
limit $1\equiv 2$. Furthermore, they satisfy the following sum rule:
\begin{equation}
\mathcal{I}_{4}\left( \mathcal{Q}_{1}+\mathcal{Q}_{2}\right) =\mathbf{I}%
_{+2}+4\mathbf{I}_{+1}+6\mathbf{I}_{0}+4\mathbf{I}_{-1}+\mathbf{I}_{-2}.
\label{sum-rule}
\end{equation}
Moreover, under the center exchange $1\leftrightarrow 2$, the polynomial $%
\mathbf{I}_{0}$ gets unchanged, whereas
\begin{equation}
\mathbf{I}_{+2}\leftrightarrow \mathbf{I}_{-2},~\mathbf{I}%
_{+1}\leftrightarrow \mathbf{I}_{-1}.  \label{prop-transf}
\end{equation}

One can compute also the following $SL_{h}\left( 2,\mathbb{R}\right) $%
-singlets:
\begin{eqnarray}
\text{Tr}\left( \frak{I}^{2}\right) &=&\mathbf{I}_{+2}\mathbf{I}_{-2}+3%
\mathbf{I}_{0}^{2}-4\mathbf{I}_{+1}\mathbf{I}_{-1};  \label{Tr(I^2)} \\
\text{Tr}\left( \frak{I}^{3}\right) &=&\mathbf{I}_{0}^{3}+\mathbf{I}_{+2}%
\mathbf{I}_{-1}^{2}+\mathbf{I}_{-2}\mathbf{I}_{+1}^{2}-\mathbf{I}_{+2}%
\mathbf{I}_{-2}\mathbf{I}_{0}-2\mathbf{I}_{+1}\mathbf{I}_{0}\mathbf{I}_{-1}.
\label{Tr(I^3)}
\end{eqnarray}
(\ref{Tr(I^2)}) and (\ref{Tr(I^3)}) are the only independent $SL_{h}\left( 2,%
\mathbb{R}\right) $-singlets which can be built out of the $3\times 3$
symmetric matrix $\frak{I}$ defined in (\ref{InvVec}), due to its very
tracelessness. Furthermore, they both vanish in the $1$-center limit $%
1\equiv 2$.

Also the polynomial $\mathbf{I}_{6}$ defined by (\ref{sextic}) is a singlet
of the horizontal symmetry $SL_{h}\left( 2,\mathbb{R}\right) $; it is
antisymmetric under $1\leftrightarrow 2$, and it vanishes when $1\equiv 2$.

The very same properties are shared by the quadratic invariant given by the
symplectic product ($\Omega $ denoting here the $Sp\left( 8,\mathbb{R}%
\right) $ metric)
\begin{eqnarray}
\mathcal{W} &\equiv &\left\langle \mathcal{Q}_{1},\mathcal{Q}%
_{2}\right\rangle \equiv \mathcal{Q}_{1}^{T}\Omega \mathcal{Q}_{2}  \notag \\
&=&-p^{0}Q_{0}-p^{1}Q_{1}-p^{2}Q_{2}-p^{3}Q_{3}+q_{0}P^{0}+q_{1}P^{1}+q_{2}P^{2}+q_{3}P^{3},
\label{SymplProd-stu}
\end{eqnarray}
which is nothing but the $\eta $-trace of the antisymmetric $\mathbb{T}$%
-tensor defined by (\ref{Ta}):
\begin{equation}
\mathcal{W}=\eta ^{\Lambda \Sigma }\mathbb{T}_{a,\Lambda \Sigma }\left(
\mathcal{Q}_{1}\mathcal{Q}_{2}\right) \equiv \text{Tr}\left( \mathbb{T}%
_{a}\right) .  \label{W-call}
\end{equation}
Thus, $\mathcal{W}$ is an $SL_{h}\left( 2,\mathbb{R}\right) $-singlet,
antisymmetric under $1\leftrightarrow 2$.

By recalling (\ref{T-bold}) and (\ref{W-call}), a particular combination of $%
SL_{h}(2,\mathbb{R})$-singlets (symmetric under $1\leftrightarrow 2$) which
will be relevant in the subsequent treatment can be defined as follows:
\begin{equation}
\mathcal{X}\equiv 2\left\| \mathbb{T}\right\| ^{2}-\frac{1}{2}\text{Tr}%
^{2}\left( \mathbb{T}_{a}\right) =2\left( \mathbf{I}^{\prime }-\mathbf{I}%
^{\prime \prime }\right) -\frac{1}{2}\mathcal{W}^{2}.  \label{x-Frak}
\end{equation}
Note that both (\ref{SymplProd-stu}) and (\ref{x-Frak}) vanish when $1\equiv
2$.\medskip

An equivalent group theoretical characterization of the quartic invariants $%
\mathbf{I}_{+2}$, $\mathbf{I}_{+1}$, $\mathbf{I}_{0}$, $\mathbf{I}_{-1}$ and
$\mathbf{I}_{-2}$ fit them into a rank-$4$ completely symmetric tensor of
the fundamental irrepr. $\mathbf{2}$ of $SL_{h}\left( 2,\mathbb{R}\right) $
itself.

This interpretation enjoys an immediate generalisation to the case of $p$
centers. Indeed, as mentioned in Sec. \ref{Hor-Symm}, in this case the
``horizontal'' combinatorics symmetry group is $SL_{h}\left( p,\mathbb{R}%
\right) $.

As a consequence, the quartic polynomial $G_{4}$-invariants which can be
obtained by computing $\mathcal{I}_{4}\left( \sum_{a=1}^{p}\mathcal{Q}%
_{a}\right) $ sit in the rank-$4$ completely symmetric tensor product of the
fundamental irrepr. $\mathbf{p}$ of $SL_{h}\left( p,\mathbb{R}\right) $, and
their number is thus given by $\binom{p+3}{4}$, which yields $1$ for $p=1$
(namely, $\mathcal{I}_{4}\left( \mathcal{Q}\right) $), $5$ for $p=2$, $15$
for $p=3$, \textit{etc}.

Furthermore, the quadratic polynomial $G_{4}$-invariants (antisymmetric
under $1\leftrightarrow 2$) sit in the rank-$2$ antisymmetric tensor product
of the fundamental irrepr. $\mathbf{2}$ of $SL_{h}\left( p,\mathbb{R}\right)
$, and their number is thus given by $\frac{p\left( p-1\right) }{2}$, which
yields $0$ for $p=1$, $1$ for $p=2$ (namely, the symplectic product $%
\mathcal{W}\equiv \left\langle \mathcal{Q}_{1},\mathcal{Q}_{2}\right\rangle $%
), $3$ for $p=3$ (namely, the three symplectic products $\mathcal{W}%
_{1}\equiv \left\langle \mathcal{Q}_{1},\mathcal{Q}_{2}\right\rangle $, $%
\mathcal{W}_{2}\equiv \left\langle \mathcal{Q}_{1},\mathcal{Q}%
_{3}\right\rangle $ and $\mathcal{W}_{3}\equiv \left\langle \mathcal{Q}_{2},%
\mathcal{Q}_{3}\right\rangle $), \textit{et cetera}. The very same holds for
the sextic $G_{4}$-polynomial invariant $\mathbf{I}_{6}$ defined by (\ref
{sextic}).

\section{\label{stu-Descendants}Independent Invariants, Constraints\newline
and their $st^{2}$ and $t^{3}$ Descendants}

We now face the issue of the independence of the various $G_{4}$-invariants
introduced so far, namely $\mathbf{I}_{+2}$, $\mathbf{I}_{+1}$, $\mathbf{I}%
^{\prime }$, $\mathbf{I}^{\prime \prime }$, $\mathbf{I}_{-1}$, $\mathbf{I}%
_{-2}$, $\mathcal{W}$ and $\mathbf{I}_{6}$, which is directly related to the
explicit derivation of the various constraints among them.

Generally, an effective method to check the functional relations (if any)
holding within a given set of $G_{4}$-invariants is the one based on the
analysis of the Jacobian matrix. In the case under consideration, one
defines the rectangular $8\times 16$ Jacobian matrix $\mathbf{J}$
\begin{equation}
\mathbf{J}\equiv \frac{\partial \mathtt{I}}{\partial \mathbf{Q}{^{\alpha }}},
\label{Jac}
\end{equation}
where
\begin{equation}
\mathtt{I}\equiv \left( \mathbf{I}_{+2},\mathbf{I}_{+1},\mathbf{I}^{\prime },%
\mathbf{I}^{\prime \prime },\mathbf{I}_{-1},\mathbf{I}_{-2},\mathcal{W},%
\mathbf{I}_{6}\right) ,  \label{I-stu}
\end{equation}
and ($\alpha =1,...,16$)
\begin{equation}
\mathbf{Q}^{\alpha }\equiv \left( \mathcal{Q}_{1}^{T},\mathcal{Q}%
_{2}^{T}\right) ^{T}=\left(
p^{0},p^{1},p^{2},p^{3},q_{0},q_{1},q_{2},q_{3},P^{0},P^{1},P^{2},P^{3},Q_{0},Q_{1},Q_{2},Q_{3}\right) ^{T}
\label{Qbold-alpha-stu}
\end{equation}
is the charge vector spanning the $16$-dimensional real vector space
\begin{equation}
V\equiv V_{1}\oplus V_{2},  \label{V-whole}
\end{equation}
where $V_{i}$ is the $8$-dimensional irrepr. space of the $\left( \mathbf{%
2,2,2}\right) $ (spin $s=\left( \frac{1}{2},\frac{1}{2},\frac{1}{2}\right) $%
) of the $U$-duality group $\ \left[ SL\left( 2,\mathbb{R}\right) \right]
^{3}$, in which the magnetic and electric charges of the BH at center $i=1,2$
sit.

By direct computation, one can check that the rank of the matrix $\mathbf{J}$
is seven; in other words, all minors of rank eight of $\mathbf{J}$ do
vanish, whereas all minors of order seven are non-zero.

A first way to explain the rank seven of $\mathbf{J}$ is as follows.

The whole vector space spanned by the charge vector $\mathbf{Q}^{\alpha }$ (%
\ref{Qbold-alpha-stu}) of the two BH centers in the $stu$ model is given by
the $16$-dimensional space $V$ defined in (\ref{V-whole}). On the other
hand, the generic (BPS) orbit of $\mathbf{Q}^{\alpha }$ is given by $%
\mathcal{O}=SL\left( 2,\mathbb{R}\right) \times SO\left( 2,2\right) $
itself, and thus it is $9$-dimensional. Thus, the general formul\ae\ (\ref
{general-1})-(\ref{charge-orbit-p}) yield that
\begin{equation}
\text{dim}_{\mathbb{R}}V=\text{dim}_{\mathbb{R}}\mathcal{O}+I_{p=2},
\label{thumb-rule}
\end{equation}
where $V$ is spanned by the multi-center charge vector $\mathbf{Q}$
belonging to the multi-center orbit $\mathcal{O}$. Thus, in the
$stu$ model the number of polynomial invariants is $I_{p=2}=16-9=7$,
in agreement with the computations reported above. One can also
check that (\ref{general-1}) applied to the $1$-center case of $stu$
model trivially yields the correct result, namely $I_{p=1}=8-7=1$
\textit{(i.e.}, the quartic invariant (\ref {I4-stu}) - in ``special
coordinates'' basis or, equivalently (\ref{I4-CV}) - in
Calabi-Vesentini basis).

As we will prove in App. \ref{App-stu}, a polynomial constraint of order $12$
relates the eight $\ \left[ SL\left( 2,\mathbb{R}\right) \right] ^{3}$%
-invariant polynomials introduced so far, namely:
\begin{equation}
\mathcal{P}_{12,stu}\equiv \mathbf{I}_{6}^{2}+\mathcal{W}\,\mathcal{X}\,%
\mathbf{I}_{6}+\text{Tr}(\frak{I}^{3})+\frac{\text{Tr}(\frak{I}^{2})\,%
\mathcal{W}^{2}}{12}-\frac{\text{Tr}(\frak{I}^{2})\,\mathcal{X}}{3}-\frac{%
\mathcal{W}^{6}}{432}+\frac{\mathcal{W}^{4}\,\mathcal{X}}{36}+\frac{5\,%
\mathcal{W}^{2}\,\mathcal{X}^{2}}{36}+\frac{4\,\mathcal{X}^{3}}{27}=0
\label{P12=0-2}
\end{equation}
This manifestly $SL_{h}\left( 2,\mathbb{R}\right) $-invariant polynomial
constraint makes the counting of independent $G_{4}$-invariant polynomials
perfectly consistent with the result and analysis presented above. Namely,
in the $stu$ model, the eight $\left[ SL\left( 2,\mathbb{R}\right) \right]
^{3}$-invariant polynomials $\mathbf{I}_{+2}$, $\mathbf{I}_{+1}$, $\mathbf{I}%
^{\prime }$, $\mathbf{I}^{\prime \prime }$, $\mathbf{I}_{-1}$, $\mathbf{I}%
_{-2}$, $\mathcal{W}$ and $\mathbf{I}_{6}$ are constrained by the $12$%
-degree relation (\ref{P12=0-2}). Thus, the number of $2$-center independent
$\left[ SL\left( 2,\mathbb{R}\right) \right] ^{3}$-invariant polynomials in
the $stu$ model is $I_{p=2}=8-1=7$, in agreement with the result (both from
Jacobian analysis and general counting) discussed above.

As discussed in Secs. \ref{Gen-Analysis-(1/2)st^2} and \ref{Gen-Analysis-t^3}
(as well as in App. \ref{App-stu}), the further reduction of the constraint (%
\ref{P12=0-2}) to the $st^{2}$ and $t^{3}$ models give rise to an hierarchy
of manifestly $SL_{h}\left( 2,\mathbb{R}\right) $-invariant polynomial
relations among the various $G_{4}$-invariants.

\section{\label{Gen-Analysis-(1/2)st^2}The $st^{2}$ Model}

Through a suitable reduction procedure (see App. \ref{App-stu}, as well as
Sec. 5 of \cite{BMOS-1}), the $stu$ model gives rise to the so-called $%
\mathcal{N}=2$, $d=4$ $st^{2}$ model. In the ``special coordinates''
symplectic frame (see \textit{e.g.} \cite{CDF-rev} and Refs. therein), this
model is defined by the prepotential
\begin{eqnarray}
F\left( X\right) &\equiv &\frac{1}{3!}d_{ijk}\frac{X^{i}X^{j}X^{k}}{X^{0}}=%
\frac{X^{1}\left( X^{2}\right) ^{2}}{X^{0}}=\left( X^{0}\right)
^{2}st^{2}\Leftrightarrow d_{122}=2;  \label{F-def-(1/2)st^2} \\
s &\equiv &\frac{X^{1}}{X^{0}},~t\equiv \frac{X^{2}}{X^{0}}.  \label{s-t-def}
\end{eqnarray}
The $Sp\left( 6,\mathbb{R}\right) $-vector of holomorphic symplectic
sections can thus be written as follows:
\begin{equation}
\mathbf{V}\equiv \left(
\begin{array}{c}
X^{0} \\
X^{1} \\
X^{2} \\
F_{0} \\
F_{1} \\
F_{2}
\end{array}
\right) =\left(
\begin{array}{c}
1 \\
s \\
t \\
-st^{2} \\
t^{2} \\
2st
\end{array}
\right) X^{0}=\left(
\begin{array}{c}
1 \\
s \\
t \\
-\mathcal{F} \\
\mathcal{F}_{1} \\
\mathcal{F}_{2}
\end{array}
\right) X^{0}.  \label{s-t-sympl-sects}
\end{equation}
Here we will not report a detailed treatment of the $st^{2}$ model (we
address the reader \textit{e.g.} to \cite{BMOS-1,stu-unveiled,CDFY-2}), we
will just confine ourselves to some basics, useful for the developments
given below.

The $st^{2}$ model is the unique example of $d$-SK geometry with dim$_{%
\mathbb{C}}=2$ (corresponding to $n_{V}=2$ vector multiplets). It is based
on the rank-$2$ factorised symmetric coset
\begin{equation}
\frac{G_{4}}{H_{4}}=\frac{SL\left( 2,\mathbb{R}\right) }{U(1)}\times \frac{%
SO\left( 2,1\right) }{SO\left( 2\right) }\sim \left[ \frac{SL\left( 2,%
\mathbb{R}\right) }{U(1)}\right] ^{2},
\end{equation}
where $G_{4}=\left[ SL\left( 2,\mathbb{R}\right) \right] ^{2}$ is the $d=4$ $%
U$-duality group, and $H_{4}=\left[ U\left( 1\right) \right] ^{2}$
its \textit{mcs}. This coset, with constant curvature $-3$
\cite{CVP}, is the first element ($n=1$) of the infinite sequence of
reducible SK
symmetric cosets $\frac{SL\left( 2,\mathbb{R}\right) }{U(1)}\times \frac{%
SO\left( 2,n\right) }{SO\left( 2\right) \times SO\left( n\right) }$ (the
so-called Jordan symmetric sequence; see \textit{e.g.} \cite{CFG,dWVVP}, and
Refs. therein).

As the $stu$ model, the $st^{2}$ model admits all classes of extremal BH
attractors \cite{AM-Refs} (for a general analysis and the treatment of
attractor-supporting charge orbits, see \textit{e.g.} \cite{BFGM1}). The BPS
solutions were known after \cite{BKRSW,Shmakova}, whereas the explicit
expression of the non-BPS $Z_{H}=0$ attractors have been obtained in \cite
{BMOS-1}. The non-BPS $Z_{H}\neq 0$ attractor solutions can also be
obtained, through a \textit{``}$stu\rightarrow st^{2}$\textit{\ reduction''}
procedure (see \textit{e.g.} Sect. 5 of \cite{BMOS-1}), by performing the
(near-)horizon limit ($\tau \rightarrow -\infty $) of the general
expressions of the $\frac{1}{2}$-BPS and non-BPS $Z_{H}\neq 0$ attractor
flows of the $stu$ model, obtained in full generality in \cite{stu-unveiled}
(see also Refs. therein, as well as \cite{CDFY-2}).

By introducing the $Sp\left( 6,\mathbb{R}\right) $-vector of charges
\begin{equation}
\mathcal{Q}\equiv \left( p^{0},p^{1},p^{2},q_{0},q_{1},q_{2}\right) ^{T},
\end{equation}
in the ``special coordinate basis'' the unique polynomial invariant
(homogeneous and quartic in the charges) of the $\left( \mathbf{2,3}\right) $
(namely spin $s=\left( \frac{1}{2},1\right) $) of the $U$-duality group $%
\left[ SL\left( 2,\mathbb{R}\right) \right] ^{2}$ reads %
\begin{equation}
\mathcal{I}_{4}\left( \mathcal{Q}\right) =-\left( p^{0}\right)
^{2}q_{0}^{2}-\left( p^{1}\right)
^{2}q_{1}^{2}-2p^{0}q_{0}p^{1}q_{1}-2p^{0}q_{0}p^{2}q_{2}+2p^{1}q_{1}p^{2}q_{2}+4q_{0}p^{1}\left( p^{2}\right) ^{2}-p^{0}q_{1}q_{2}^{2},
\label{I4-(1/2)st^2}
\end{equation}
which can be obtained from (\ref{I4}) by specifying $d_{122}=2$ and $%
d^{122}=1/2$, consistently with the non-linear relation (\ref{symmetric-cond}%
).

By considering $2$-center extremal BHs in the $st^{2}$ model, with the
charge vectors associated to the two centers respectively reading
\begin{eqnarray}
\mathcal{Q}_{1} &\equiv &\left( p^{0},p^{1},p^{2},q_{0},q_{1},q_{2}\right)
^{T};  \label{Q1-st^2} \\
\mathcal{Q}_{2} &\equiv &\left( P^{0},P^{1},P^{2},Q_{0},Q_{1},Q_{2}\right)
^{T},  \label{Q2-st^2}
\end{eqnarray}
the eight $G_{4}$-invariant polynomials $\mathbf{I}_{+2}$, $\mathbf{I}_{+1}$%
, $\mathbf{I}^{\prime }$, $\mathbf{I}^{\prime \prime }$, $\mathbf{I}_{-1}$, $%
\mathbf{I}_{-2}$, $\mathcal{W}$ and $\mathbf{I}_{6}$ can be obtained from
their very definitions (\ref{I+2-CV})-(\ref{I-2-CV}), (\ref{SymplProd-stu})
and (\ref{sextic}) by simply specifying $\Lambda =0,1,2$ (and thus using the
metric $\eta _{\Lambda \Sigma }=\eta ^{\Lambda \Sigma }$ of $SO\left(
2,1\right) $ to raise and lower the indices).

In order to establish the independence of such $2$-center $\left[ SL\left( 2,%
\mathbb{R}\right) \right] ^{2}$-invariants introduced above, we will exploit
the Jacobian method used above for the $stu$ model, adapted to the model
under consideration. To this end, one defines the rectangular $6\times 12$
Jacobian matrix $\mathbf{J}$ (\ref{Jac}), where $\mathtt{I}$ is defined in (%
\ref{I-stu}),with ($\alpha =1,...,12$)
\begin{equation}
\mathbf{Q}^{\alpha }\equiv \left( \mathcal{Q}_{1}^{T},\mathcal{Q}%
_{2}^{T}\right) ^{T}=\left(
p^{0},p^{1},p^{2},q_{0},q_{1},q_{2},P^{0},P^{1},P^{2},Q_{0},Q_{1},Q_{2}%
\right) ^{T}  \label{Qbold-alpha-st^2}
\end{equation}
is the charge vector spanning the $12$-dimensional vector space $V$ given by
(\ref{V-whole}), where now $V_{i}$ is the $6$-dimensional irrepr. space of
the $\left( \mathbf{2,3}\right) $ (spin $s=\left( \frac{1}{2},1\right) $) of
the $U$-duality group$\ \left[ SL\left( 2,\mathbb{R}\right) \right] ^{2}$,
in which the magnetic and electric charges of the BH at center $i=1,2$ sit.

By direct computation, one can check that the rank of the matrix $\mathbf{J}$
for the $st^{2}$ model is six; in other words, all minors of rank six of $%
\mathbf{J}$ are non-zero, whereas all minors of rank seven and eight do
vanish.

Similarly to the discussion done for the $stu$ model, a simple venue for the
explanation for the rank six of $\mathbf{J}$ in the $st^{2}$ model is as
follows.

The whole vector space spanned by the charge vector $\mathbf{Q}^{\alpha }$ (%
\ref{Qbold-alpha-st^2}) of the two BH centers in the $st^{2}$ model is given
by the $12$-dimensional space $V$ defined in (\ref{V-whole}). On the other
hand, the generic (BPS) orbit of $\mathbf{Q}^{\alpha }$ is given by $%
\mathcal{O}=SL\left( 2,\mathbb{R}\right) \times SO\left( 2,1\right) $
itself, and thus it is $6$-dimensional. Thus, by applying the relation (\ref
{general-1}) (holding in the theory of polynomial invariants of Lie groups)
to the $st^{2}$ model, the final result on the number of polynomial
invariants is $I_{p=2}=12-6=6$, in agreement with the computations reported
above. One can also check that (\ref{general-1}) applied to the $1$-center
case of $st^{2}$ model trivially yields the correct result, namely $%
I_{p=1}=6-5=1$ \textit{(i.e.}, the quartic invariant
(\ref{I4-(1/2)st^2}) - in ``special coordinates'' basis or,
equivalently (\ref{I4-CV}) - in Calabi-Vesentini basis).\medskip

The above counting of independent $2$-center polynomial invariants of the $U$%
-duality group $\left[ SL\left( 2,\mathbb{R}\right) \right] ^{2}$ of the $%
st^{2}$ model is consistent with the number of independent, manifestly $%
SL_{h}\left( 2,\mathbb{R}\right) $-invariant polynomial relations holding
for the $st^{2}$ model itself.

Indeed, as we will detail in App. \ref{App-stu}, starting from the $stu$
model and its constraint (\ref{P12=0-2}), a suitable reduction to $st^{2}$
model determines the following two manifestly $SL_{h}\left( 2,\mathbb{R}%
\right) $-invariant constraints:
\begin{eqnarray}
\mathcal{P}_{16,st^{2}} &\equiv &\mathcal{P}_{16,t^{3}}-\frac{32}{3}\text{Tr}%
\left( \frak{I}^{2}\right) \mathcal{X}^{2}+\frac{8}{9}\mathcal{W}^{4}%
\mathcal{X}^{2}+\frac{64}{27}\mathcal{W}^{2}\mathcal{X}^{3}+\frac{16}{9}%
\mathcal{X}^{4}=0;  \label{P16=0-2} \\
\mathcal{P}_{8,st^{2}} &\equiv &-12\,\text{Tr}\left( \frak{I}^{2}\right)
+24\,\mathbf{I}_{6}\,\mathcal{W}+(\mathcal{W}^{2}+2\mathcal{X})^{2}=0,
\label{P8=0-2}
\end{eqnarray}
where
\begin{equation}
\mathcal{P}_{16,t^{3}}\equiv 16\text{Tr}^{2}\left( \frak{I}^{2}\right) +64%
\text{Tr}\left( \frak{I}^{3}\right) \mathcal{W}^{2}+\frac{8}{3}\text{Tr}%
\left( \frak{I}^{2}\right) \mathcal{W}^{4}-\frac{1}{27}\mathcal{W}^{8}.
\label{P16-t^3}
\end{equation}

Note that (\ref{P8=0-2}) expresses $\mathbf{I}_{6}$ in terms of the other
invariants, whereas (\ref{P16=0-2}) is the constraint which decreases the
number of independent polynomial invariants from seven to six.

It is also worth pointing out that the very structure of constraints (\ref
{P12=0-2}) and (\ref{P16-t^3}) is determined by the underlying $SL_{h}\left(
2,\mathbb{R}\right) $-invariance; for instance, this latter constrains the
inhomogeneous term of (\ref{P16-t^3}) to be the square of the coefficient of
$\mathcal{W}^{4}$ in the same equation. Also, the fact that a term
proportional to $\mathcal{W}^{6}$ is missing in Eq. (\ref{P16-t^3}) is due
to the tracelessness of $\frak{I}$ itself (recall (\ref{InvVec})): Tr$\frak{I%
}=0$.

The manifestly $SL_{h}\left( 2,\mathbb{R}\right) $-invariant polynomial
constraints (\ref{P8=0-2})-(\ref{P16=0-2}) make the counting of independent $%
G_{4}$-invariant polynomials perfectly consistent with the result and
analysis presented above. Namely, in the $st^{2}$ model, the eight $\left[
SL\left( 2,\mathbb{R}\right) \right] ^{2}$-invariant polynomials $\mathbf{I}%
_{+2}$, $\mathbf{I}_{+1}$, $\mathbf{I}^{\prime }$, $\mathbf{I}^{\prime
\prime }$, $\mathbf{I}_{-1}$, $\mathbf{I}_{-2}$, $\mathcal{W}$ and $\mathbf{I%
}_{6}$ are constrained by the $8$-degree and $16$-degree relations
respectively given by Eqs. (\ref{P8=0-2}) and (\ref{P16=0-2}). Thus, the
number of $2$-center independent $\left[ SL\left( 2,\mathbb{R}\right) \right]
^{2}$-invariant polynomials in the $st^{2}$ model is $I_{p=2}=8-2=6$, in
agreement with the result (both from Jacobian analysis and general counting)
discussed above.

\section{\label{Gen-Analysis-t^3}The $t^{3}$ Model}

Through a suitable reduction procedure (see App. \ref{App-stu}, as well as
Sec. 5 of \cite{BMOS-1}), the $stu$ model gives rise to the so-called $%
\mathcal{N}=2$, $d=4$ $t^{3}$ model. In the ``special coordinates''
symplectic frame (see \textit{e.g.} \cite{CDF-rev} and Refs. therein), this
model is defined by the prepotential
\begin{eqnarray}
F\left( X\right) &\equiv &\frac{1}{3!}d_{ijk}\frac{X^{i}X^{j}X^{k}}{X^{0}}=%
\frac{\left( X^{1}\right) ^{3}}{X^{0}}=\left( X^{0}\right)
^{2}t^{3}\Leftrightarrow d_{111}=6;  \label{F-def} \\
t &\equiv &\frac{X^{1}}{X^{0}}.  \label{t-def}
\end{eqnarray}

It is worth recalling that the $t^{3}$ model is the unique example of $d$%
-special K\"{a}hler (SK) geometry \cite{dWVVP} with dim$_{\mathbb{C}}=1$
(corresponding to $n_{V}=1$ vector multiplet). It is based on the rank-$1$
symmetric coset
\begin{equation}
\frac{G_{4}}{H_{4}}=\frac{SL\left( 2,\mathbb{R}\right) }{U(1)},
\end{equation}
where $G_{4}=SL\left( 2,\mathbb{R}\right) $ is the $d=4$ $U$-duality group,
and $H_{4}=U\left( 1\right) $ its maximal compact subgroup (\textit{mcs}).
This coset, with constant curvature $-\frac{2}{3}$ \cite{CVP}, is an
isolated case within the classification of homogeneous symmetric non-compact
SK manifolds (see \textit{e.g.} \cite{CFG,dWVVP}, and Refs. therein).

Through the definition (\ref{t-def}) of the \textit{projective} coordinate $%
t $, the $Sp\left( 4,\mathbb{R}\right) $-vector of holomorphic symplectic
sections can thus be written as follows:
\begin{equation}
\mathbf{V}\equiv \left(
\begin{array}{c}
X^{0} \\
X^{1} \\
F_{0} \\
F_{1}
\end{array}
\right) =\left(
\begin{array}{c}
1 \\
t \\
-t^{3} \\
3t^{2}
\end{array}
\right) X^{0}=\left(
\begin{array}{c}
1 \\
t \\
-\mathcal{F} \\
\mathcal{F}_{1}
\end{array}
\right) X^{0}.  \label{sympl-sects}
\end{equation}

By introducing the $Sp\left( 4,\mathbb{R}\right) $-vector of magnetic and
electric charges
\begin{equation}
\mathcal{Q}\equiv \left( p^{0},p,q_{0},q\right) ^{T},  \label{Q}
\end{equation}
the unique invariant (homogeneous quartic polynomial in the charges) of the $%
\mathbf{4}$ (spin $s=3/2$) irrepr. of the $U$-duality group $SL\left( 2,%
\mathbb{R}\right) $ reads (in the ``special coordinates'' symplectic frame)%
%
%
%
%
%
%
%
%
%
%
%
%
%
%
%
%
%
%
%
%
%
%
%
%
%
%
%
%
%
%
%
%
%
%
%
%
%
%
%
%
\begin{equation}
\mathcal{I}_{4}\left( \mathcal{Q}\right) \equiv -\left( p^{0}\right)
^{2}q_{0}^{2}+\frac{1}{3}p^{2}q^{2}-2p^{0}q_{0}pq+4q_{0}p^{3}-\frac{4}{27}%
p^{0}q^{3}.  \label{I4-t^3}
\end{equation}
In general, the sign of $\mathcal{I}_{4}$ is related to the supersymmetry
properties of the only two classes of extremal BH attractors \cite{AM-Refs}
exhibited by the $t^{3}$ model: namely, $\mathcal{I}_{4}>0$ and $\mathcal{I}%
_{4}<0$ for $\frac{1}{2}$-BPS and non-BPS $Z_{H}\neq 0$ attractor solutions,
respectively (see \textit{e.g.} Appendix II of \cite{BFGM1}). The $t^{3}$
model has been the first supergravity model whose Attractor Eqs. have been
completely solved. The BPS attractor solution were known after \cite
{BKRSW,Shmakova}, and in \cite{Saraikin-Vafa-1} also the non-BPS $Z_{H}\neq
0 $ attractor solutions were completely determined (see also \cite{TT1}). It
is worth here pointing out that these results can also be obtained, through
a \textit{``}$stu\rightarrow st^{2}\rightarrow t^{3}$\textit{\ reduction''}
procedure (see \textit{e.g.} Sect. 5 of \cite{BMOS-1}), by performing the
(near-)horizon limit ($\tau \rightarrow -\infty $) of the general
expressions of the $\frac{1}{2}$-BPS and non-BPS $Z_{H}\neq 0$ attractor
flows\footnote{%
Through the \textit{``}$stu\rightarrow st^{2}\rightarrow t^{3}$\textit{\
reduction''} procedure, the non-BPS $Z_{H}=0$ attractor flow of $stu$ model
consistently degenerates into the $\frac{1}{2}$-BPS attractor flow of the $%
t^{3}$ model.} of the $stu$ model, obtained in full generality in \cite
{stu-unveiled} (see also Refs. therein, as well as \cite{CDFY-1}).

Note that (\ref{I4-t^3}) can be obtained from the general formula (\ref{I4})
by specifying $d_{111}=6$ and $d^{111}=2/9$, consistently with the
non-linear relation (\ref{symmetric-cond}).

Let us now consider $2$-center extremal BHs in the $t^{3}$ model, with the
charge vectors associated to the two centers respectively reading
\begin{eqnarray}
\mathcal{Q}_{1} &\equiv &\left( p^{0},p,q_{0},q\right) ^{T};  \label{Q1} \\
\mathcal{Q}_{2} &\equiv &\left( P^{0},P,Q_{0},Q\right) ^{T}.  \label{Q2}
\end{eqnarray}
By working in the ``special coordinates'' symplectic frame, one can write
down the components of the $\mathbf{5}$ (spin $s=2$) irrepr. of the
horizontal symmetry $SL_{h}\left( 2,\mathbb{R}\right) $ by recalling Eqs. (%
\ref{sum-rule}) and (\ref{I4-t^3}). Then, it is immediate to obtain the
following expressions:
\begin{eqnarray}
\mathbf{I}_{+2}\left( \mathcal{Q}_{1}^{4}\right) &\equiv &-\left(
p^{0}\right) ^{2}q_{0}^{2}+\frac{1}{3}p^{2}q^{2}-2p^{0}q_{0}pq+4q_{0}p^{3}-%
\frac{4}{27}p^{0}q^{3};  \label{I+2} \\
&&  \notag \\
\mathbf{I}_{+1}\left( \mathcal{Q}_{1}^{3}\mathcal{Q}_{2}\right) &\equiv &-%
\frac{1}{2}\left[ \left( p^{0}\right) ^{2}q_{0}Q_{0}+p^{0}q_{0}^{2}P^{0}%
\right] +\frac{1}{6}\left( p^{2}qQ+pq^{2}P\right) +3p^{2}q_{0}P+p^{3}Q_{0}
\notag \\
&&-\frac{1}{9}p^{0}q^{2}Q-\frac{1}{27}q^{3}P^{0}-\frac{1}{2}\left(
p^{0}q_{0}pQ+p^{0}q_{0}qP+p^{0}pqQ_{0}+pq_{0}qP^{0}\right) ;  \label{I+1} \\
&&  \notag \\
\mathbf{I}_{0}\left( \mathcal{Q}_{1}^{2}\mathcal{Q}_{2}^{2}\right) &\equiv &-%
\frac{1}{6}\left[ \left( p^{0}\right) ^{2}Q_{0}^{2}+q_{0}^{2}\left(
P^{0}\right) ^{2}\right] -\frac{2}{3}p^{0}q_{0}P^{0}Q_{0}+\frac{1}{18}\left(
p^{2}Q^{2}+q^{2}P^{2}\right) +\frac{2}{9}pqPQ  \notag \\
&&+2\left( pq_{0}P^{2}+p^{2}PQ_{0}\right) -\frac{2}{27}\left(
p^{0}qQ^{2}+q^{2}P^{0}Q\right)  \notag \\
&&-\frac{1}{3}\left(
p^{0}q_{0}PQ+pqP^{0}Q_{0}+p^{0}pQ_{0}Q+q_{0}qP^{0}P+p^{0}qPQ_{0}+pq_{0}P^{0}Q\right) ;
\label{I0-t^3} \\
&&  \notag \\
\mathbf{I}_{-1}\left( \mathcal{Q}_{1}\mathcal{Q}_{2}^{3}\right) &\equiv &-%
\frac{1}{2}\left[ q_{0}\left( P^{0}\right) ^{2}Q_{0}+p^{0}P^{0}Q_{0}^{2}%
\right] +\frac{1}{6}\left( qP^{2}Q+pPQ^{2}\right) +3pP^{2}Q_{0}+q_{0}P^{3}
\notag \\
&&-\frac{1}{9}qP^{0}Q^{2}-\frac{1}{27}p^{0}Q^{3}-\frac{1}{2}\left(
qP^{0}Q_{0}P+pP^{0}Q_{0}Q+q_{0}P^{0}PQ+p^{0}PQ_{0}Q\right) ;  \label{I-1} \\
&&  \notag \\
\mathbf{I}_{-2}\left( \mathcal{Q}_{2}^{4}\right) &=&-\left( P^{0}\right)
^{2}Q_{0}^{2}+\frac{1}{3}P^{2}Q^{2}-2P^{0}Q_{0}PQ+4Q_{0}P^{3}-\frac{4}{27}%
P^{0}Q^{3}.  \label{I-2}
\end{eqnarray}

In order to establish the independence of such $2$-center $SL\left( 2,%
\mathbb{R}\right) $-invariants, we will exploit the Jacobian method used
above for the $stu$ and $st^{2}$ models. To this end, one defines the
rectangular $6\times 8$ Jacobian matrix $\mathbf{J}$ (\ref{Jac}), where $%
\mathtt{I}$ is defined by
\begin{equation}
\mathtt{I}\equiv \left( \mathbf{I}_{+2},\mathbf{I}_{+1},\mathbf{I}_{0},%
\mathbf{I}_{-1},\mathbf{I}_{-2},\mathcal{W}\right)
\end{equation}
and ($\alpha =1,...,8$)
\begin{equation}
\mathbf{Q}^{\alpha }\equiv \left( \mathcal{Q}_{1}^{T},\mathcal{Q}%
_{2}^{T}\right) ^{T}=\left( p^{0},p,q_{0},q,P^{0},P,Q_{0},Q\right) ^{T}
\label{Q-bold-alpha-t^3}
\end{equation}
is the charge vector spanning the $8$-dimensional vector space (\ref{V-whole}%
), where $V_{i}$ is the $4$-dimensional irrepr. space of the $\mathbf{4}$
(spin $s=3/2$) irrepr. of the $U$-duality group $SL\left( 2,\mathbb{R}%
\right) $, in which the magnetic and electric charges of the BH at center $%
i=1,2$ sit.

By direct computation, one can check that the rank of the matrix $\mathbf{J}$
in the $t^{3}$ model is five. Namely, all minors of rank six of $\mathbf{J}$
do vanish, whereas all minors of rank five are non-zero.

Similarly to the discussion done for the $stu$ and $st^{2}$ models, a simple
venue for the explanation for the rank five of $\mathbf{J}$ in the $t^{3}$
model is as follows.

The whole vector space spanned by the charge vector $\mathbf{Q}^{\alpha }$ (%
\ref{Q-bold-alpha-t^3}) of the two BH centers in the $t^{3}$ model is given
by the $8$-dimensional space $V$ defined in (\ref{V-whole}). On the other
hand, the generic (BPS) orbit of $\mathbf{Q}^{\alpha }$ is given by $%
\mathcal{O}=SL\left( 2,\mathbb{R}\right) $ itself, and thus it is $3$%
-dimensional. Thus, by applying the relation (\ref{general-1}) to the $%
st^{2} $ model, the final result on the number of polynomial invariants is $%
I_{p=2}=8-3=5$, in agreement with the computations reported above. One can
also check that (\ref{general-1}) applied to the $1$-center case of $t^{3}$
model trivially yields the correct result, namely $I_{p=1}=4-3=1$ \textit{%
(i.e.}, the quartic invariant (\ref{I4-t^3}) - in ``special
coordinates'' basis or, equivalently (\ref{I4-CV}) - in
Calabi-Vesentini basis).\medskip

The above counting of independent $2$-center polynomial invariants of the $U$%
-duality group $SL\left( 2,\mathbb{R}\right) $ of the $t^{3}$ model is
consistent with the number of independent, manifestly $SL_{h}\left( 2,%
\mathbb{R}\right) $-invariant polynomial relations holding for the $t^{3}$
model itself.

Before proceeding, it should be remarked that the \textit{irreducible} (rank-%
$1$) $t^{3}$ model is \textit{sui generis} with respect to the \textit{%
reducible} $stu$ (rank-$3$) and $st^{2}$ (rank-$2$) models. Indeed,
while the Calabi-Vesentini \cite{CDFVP-1} $\mathbb{T}$-tensor
formalism introduced above can be applied to both the $stu$ and
$st^{2}$, it requires dome further modifications in order to be
applied to the $t^{3}$ model. This can essentially be traced back to
the fact that, while the $stu$ and $st^{2}$ models are the first two
elements ($n=2$ and $n=1$, respectively) of the aforementioned
$\mathcal{N}=2$ Jordan symmetric sequence, the $t^{3}$ model is an
isolated case within the classification of homogeneous symmetric
non-compact SK manifolds (see \textit{e.g.} \cite{CFG,dWVVP}, and
Refs.
therein). As a consequence, the consistent application of the $\mathbb{T}$%
-tensor (Calabi-Vesentini) formalism to the $t^{3}$ model requires
some \textit{ad hoc} modifications (leading to a ``constrained''
Calabi-Vesentini symplectic frame), which are derived and studied in
App. \ref{App-t^3}.

As we will detail in App. \ref{App-stu}, starting from the $stu$ model and
its constraint (\ref{P12=0-2}), a suitable reduction to the $t^{3}$ model
determines the following two manifestly $SL_{h}\left( 2,\mathbb{R}\right) $%
-invariant constraints (recall (\ref{Ta}), (\ref{W-call}) and (\ref{x-Frak}%
)):
\begin{eqnarray}
\mathcal{X} &=&0\Leftrightarrow \left\| \mathbb{T}\right\| ^{2}=\frac{1}{4}%
\mathcal{W}^{2}=\frac{1}{4}\text{Tr}^{2}\left( \mathbb{T}_{a}\right)
;
\label{t^3-rell} \\
\mathcal{P}_{8,t^{3}} &\equiv &24\,\mathbf{I}_{6}\mathcal{W}-12\text{Tr}%
\left( \frak{I}^{2}\right) +\mathcal{W}^{4}=0,  \label{P8=0-3}
\end{eqnarray}
and then (recall (\ref{P16-t^3}))
\begin{equation}
\mathcal{P}_{16,t^{3}}=0.\label{P16-t^3=0}
\end{equation}

Note that the constraints (\ref{P16=0-2}) and (\ref{P16-t^3=0}) can be seen
as a quartic algebraic equation in $\mathcal{W}^{2}$, and it can be checked
that only one real positive out of the four generally complex roots exists
in the case, thus uniquely matching the square of the symplectic product of $%
\mathcal{Q}_{1}$ and $\mathcal{Q}_{2}$.

Since $\mathbf{I}_{+2}$, $\mathbf{I}_{+1}$, $\mathbf{I}^{\prime }$, $\mathbf{%
I}^{\prime \prime }$, $\mathbf{I}_{-1}$ and $\mathbf{I}_{-2}$ all reduce to $%
\mathcal{I}_{4}\left( \mathcal{Q}\right) $, and $\mathcal{W}$ and $\mathcal{X%
}$ both vanish in the $1$-center limit $1\equiv 2$, it is immediate to check
that all above constraints identically vanish in such a limit. Also notice
that all above constraints gets greatly simplified when $\mathcal{W}=0$
(namely, for \textit{mutually local} charge vectors $\mathcal{Q}_{1}$ and $%
\mathcal{Q}_{2}$).

The manifestly $SL_{h}\left( 2,\mathbb{R}\right) $-invariant polynomial
constraints (\ref{t^3-rell})-(\ref{P16-t^3=0}) make the counting of
independent $G_{4}$-invariant polynomials perfectly consistent with the
result and analysis presented above. Namely, in the $t^{3}$ model, the eight
$SL\left( 2,\mathbb{R}\right) $-invariant polynomials $\mathbf{I}_{+2}$, $%
\mathbf{I}_{+1}$, $\mathbf{I}^{\prime }$, $\mathbf{I}^{\prime \prime }$, $%
\mathbf{I}_{-1}$, $\mathbf{I}_{-2}$, $\mathcal{W}$ and $\mathbf{I}_{6}$ are
constrained by the $4$-degree, $8$-degree and $16$-degree relations
respectively given by Eqs. (\ref{t^3-rell}), (\ref{P8=0-3}) and (\ref
{P16-t^3=0}). Thus, the number of $2$-center independent $SL\left( 2,\mathbb{%
R}\right) $-invariant polynomials in the $t^{3}$ model is $I_{p=2}=8-3=5$,
in agreement with the result (both from Jacobian analysis and general
counting) discussed above.

\section{\label{SO_h(2,2)}Extension to $SO_{h}^{v}\left( 2,2\right) $
Symmetry and the Gramian Matrix}

The treatment given in previous Secs. relies on the fact that an $%
SL_{h}\left( 2,\mathbb{R}\right) $-covariant basis is given by the quintet $%
\frak{I}$, and the two singlets $\mathcal{W}$ and $\mathcal{X}$
(respectively defined by (\ref{InvVec}), (\ref{W-call}) and (\ref{x-Frak})).
By using such a basis, the following $\left[ SL_{h}\left( 2,\mathbb{R}%
\right) \times G_{4}\right] $-invariant set of polynomials can be
constructed:
\begin{equation}
\underset{\deg =2}{\mathcal{W}},~\underset{\deg =4}{\mathcal{X}},~\underset{%
\deg =8}{\text{Tr}\left( \frak{I}^{2}\right) },~\underset{\deg =12}{\text{Tr}%
\left( \frak{I}^{3}\right) },  \label{4-1}
\end{equation}
where the degree in charges has been indicated.

However, a lower degree $\left[ SL_{h}\left( 2,\mathbb{R}\right) \times G_{4}%
\right] $-invariant polynomial, namely $\mathbf{I}_{6}$ defined by (\ref
{sextic}), is related to Tr$\left( \frak{I}^{3}\right) $ through the degree-$%
12$ polynomial constraints (\ref{P12=0-2}). Actually, if in (\ref{4-1}) Tr$%
\left( \frak{I}^{3}\right) $ is replaced by $\mathbf{I}_{6}$, one obtains
the following complete set of $\left[ SL_{h}\left( 2,\mathbb{R}\right)
\times G_{4}\right] $-invariant, with ``minimal'' degrees in charges
(indicated by subscripts):
\begin{equation}
\underset{\deg =2}{\mathcal{W}},~\underset{\deg =4}{\mathcal{X}},~\underset{%
\deg =6}{\mathbf{I}_{6}},~\underset{\deg =8}{\text{Tr}\left( \frak{I}%
^{2}\right) }.  \label{4-2}
\end{equation}
Then, by the theory of invariant polynomials of classical Lie groups, one is
guaranteed that any other higher-order invariant is related to the
lowest-degree invariants by an algebraic relation.

As given by Eqs. (\ref{TrG})-(\ref{Tr(G^4)}) below, the set (\ref{4-2}) is
naturally related to the symmetry
\begin{equation}
SO_{h}^{v}\left( 2,2\right) \equiv SL_{h}\left( 2,\mathbb{R}\right) \times
SL_{v}\left( 2,\mathbb{R}\right) \overset{\mathbb{C}}{\sim }SO_{h}^{v}\left(
4,\mathbb{C}\right) ,  \label{formula-1}
\end{equation}
which is the direct product of the ``horizontal'' group $SL_{h}\left( 2,%
\mathbb{R}\right) $ introduced in Sec. \ref{Hor-Symm} and of the $%
SL_{v}\left( 2,\mathbb{R}\right) $ factor (the upperscript ``$v$'' stands
for ``vertical'') in the $d=4$ $U$-duality group $G_{4}$ (characterising the
three models $stu$, $st^{2}$ and $t^{3}$ treated above, as well as the whole
infinite sequences (\ref{N=2-J-symm-seq}) and (\ref{N=4}); see Sec. \ref
{Gen-Reducible-N=2-N=4}). The last step in (\ref{formula-1}) denotes the
isomorphism with the complex group $SO\left( 4,\mathbb{C}\right) $. In the
following treatment, we will work with complex groups, thus $SL\left( 2,%
\mathbb{R}\right) \overset{\mathbb{C}}{\sim }SL\left( 2,\mathbb{C}\right) $
and $SO\left( 2,2\right) \overset{\mathbb{C}}{\sim }SO\left( 4,\mathbb{C}%
\right) $ and, where denoted, we will then perform the suitable Wick
rotation to get the appropriate real form.

In order to highlight the relation between the set (\ref{4-2}) and the
symmetry group $SO_{h}^{v}\left( 4,\mathbb{C}\right) $ defined in (\ref
{formula-1}), it is convenient to introduce the $4\times 4$ complex
symmetric, manifestly $SO_{h}^{v}\left( 4,\mathbb{C}\right) $-covariant
so-called \textit{Gramian matrix }\cite{Gramian}
\begin{equation}
\mathbf{G}\equiv \left(
\begin{array}{cccc}
z_{0}^{2} & z_{0}\cdot z_{1} & z_{0}\cdot z_{2} & z_{0}\cdot z_{3} \\
z_{0}\cdot z_{1} & z_{1}^{2} & z_{1}\cdot z_{2} & z_{1}\cdot z_{3} \\
z_{0}\cdot z_{2} & z_{1}\cdot z_{2} & z_{2}^{2} & z_{2}\cdot z_{3} \\
z_{0}\cdot z_{3} & z_{1}\cdot z_{3} & z_{2}\cdot z_{3} & z_{3}^{2}
\end{array}
\right) ,  \label{G-matrix}
\end{equation}
where
\begin{equation}
\left(
\begin{array}{c}
z_{0,\Lambda } \\
z_{1,\Lambda } \\
z_{2,\Lambda } \\
z_{3,\Lambda }
\end{array}
\right) \equiv \frac{1}{2}\left(
\begin{array}{cccc}
1 & 0 & 0 & 1 \\
0 & -i & -i & 0 \\
0 & 1 & -1 & 0 \\
-i & 0 & 0 & +i
\end{array}
\right) \left(
\begin{array}{c}
p_{\Lambda } \\
q_{\Lambda } \\
P_{\Lambda } \\
Q_{\Lambda }
\end{array}
\right)  \label{G-matrix-2}
\end{equation}
and the squared norms, scalar products and index raising and lowering are
defined through the suitable $SO\left( 2,n\right) $-metrics $\eta _{\Lambda
\Sigma }$ and $\eta ^{\Lambda \Sigma }$ (in $\mathcal{N}=4$ theory, $%
SO\left( 2,n\right) $ is replaced by $SO\left( 6,n\right) $; see Eq. (\ref
{N=4}) below).

Then, by denoting the eigenvalues of $\mathbf{G}$ with $\lambda
_{i}$ ($i=1,...,4$), the characteristic equation of $\mathbf{G}$
reads:
\begin{equation}
det\left( \mathbf{G}-\lambda \mathbb{I}\right)
=\prod_{i=1}^{4}\left( \lambda -\lambda _{i}\right) =\lambda
^{4}+a\lambda ^{3}+b\lambda ^{2}+c\lambda +d=0.
\label{N=8-d=4-char-eq}
\end{equation}
As proved in App. \ref{Proof-Completeness}, the characteristic
equation can be used as a generating function for manifestly
$SO_{h}^{v}\left( 4,\mathbb{C}\right) $-invariant polynomials.
Indeed, by recalling the
\textit{Newton's identities} \cite{Newton-Refs}, one can compute that ($%
a,b,c,d\in \mathbb{R}$; see also \cite{DFL-0-brane})
\begin{eqnarray}
a &\equiv &-Tr\mathbf{G}=-\left( \lambda _{1}+\lambda _{2}+\lambda
_{3}+\lambda _{4}\right) ;  \label{a} \\
&&  \notag \\
b &\equiv &\frac{1}{2}\left[ \left( Tr\mathbf{G}\right) ^{2}-Tr\left(
\mathbf{G}^{2}\right) \right] =  \notag \\
&=&\lambda _{1}\lambda _{2}+\lambda _{1}\lambda _{3}+\lambda _{1}\lambda
_{4}+\lambda _{2}\lambda _{3}+\lambda _{2}\lambda _{4}+\lambda _{3}\lambda
_{4};  \label{b} \\
&&  \notag \\
c &\equiv &-\frac{1}{6}\left[ \left( Tr\mathbf{G}\right) ^{3}+2Tr\left(
\mathbf{G}^{3}\right) -3Tr\left( \mathbf{G}^{2}\right) Tr\mathbf{G}\right] =
\notag \\
&=&-\left( \lambda _{1}\lambda _{2}\lambda _{3}+\lambda _{1}\lambda
_{2}\lambda _{4}+\lambda _{1}\lambda _{3}\lambda _{4}+\lambda _{2}\lambda
_{3}\lambda _{4}\right) ;  \label{c} \\
&&  \notag \\
d &\equiv &\frac{1}{4}\left[
\begin{array}{l}
\frac{1}{6}\left( Tr\mathbf{G}\right) ^{4}+\frac{1}{2}\left[ Tr\left(
\mathbf{G}^{2}\right) \right] ^{2}+\frac{4}{3}Tr\left( \mathbf{G}^{3}\right)
Tr\mathbf{G} \\
\\
-Tr\left( \mathbf{G}^{4}\right) -Tr\left( \mathbf{G}^{2}\right) \left( Tr%
\mathbf{G}\right) ^{2}
\end{array}
\right] =  \notag \\
&=&det\mathbf{G}=\lambda _{1}\lambda _{2}\lambda _{3}\lambda _{4}.  \notag \\
&&  \label{d}
\end{eqnarray}

By computing $Tr\mathbf{G}$, $Tr\left( \mathbf{G}^{2}\right) $, $Tr\left(
\mathbf{G}^{3}\right) $ and $Tr\left( \mathbf{G}^{4}\right) $, and then
performing the Wick rotation
\begin{equation}
\left\{
\begin{array}{l}
z_{1,\Lambda }\longrightarrow iz_{1,\Lambda }; \\
z_{3,\Lambda }\longrightarrow iz_{3,\Lambda }
\end{array}
\right.  \label{Wick-rot}
\end{equation}
in order to switch from $SO_{h}^{v}\left( 4,\mathbb{C}\right) $ back to $%
SO_{h}^{v}\left( 2,2\right) $, the following results can be achieved:
\begin{eqnarray}
Tr\mathbf{G} &=&-\mathcal{W};  \label{TrG} \\
Tr\left( \mathbf{G}^{2}\right) &=&\frac{1}{2}\left( \mathcal{W}^{2}-2%
\mathcal{X}\right) ;  \label{Tr(G^2)} \\
Tr\left( \mathbf{G}^{3}\right) &=&\frac{1}{4}\left( 6\mathbf{I}_{6}-\mathcal{%
W}^{3}+6\mathcal{WX}\right) ;  \label{Tr(G^3)} \\
Tr\left( \mathbf{G}^{4}\right) &=&-\frac{1}{48}\left[ 12\text{Tr}\left(
\frak{I}^{2}\right) +72\mathbf{I}_{6}\mathcal{W}-7\mathcal{W}^{4}+68\mathcal{%
W}^{2}\mathcal{X}-28\mathcal{X}^{2}\right] .  \label{Tr(G^4)}
\end{eqnarray}
Thus, by virtue of Eq. (\ref{d}), it follows that (recall definition (\ref
{P8=0-2}))
\begin{eqnarray}
det\mathbf{G} &=&\frac{1}{16\cdot 12}\left[ 12\text{Tr}\left( \frak{I}%
^{2}\right) -24\mathbf{I}_{6}\mathcal{W}-\left( \mathcal{W}^{2}+2\mathcal{X}%
\right) ^{2}\right] =-\frac{1}{16\cdot 12}\mathcal{P}_{8}; \\
det\left( \mathbf{G}-\lambda \mathbb{I}\right) &=&\lambda ^{4}+\mathcal{W}%
\lambda ^{3}+\left\| \mathbb{T}\right\| ^{2}\lambda ^{2}-\frac{1}{2}\mathbf{I}%
_{6}\lambda +det\mathbf{G}=0.
\end{eqnarray}

The relations (\ref{TrG})-(\ref{Tr(G^4)}) establish the connection between
the set (\ref{4-2}) and the set of manifestly $SO_{h}^{v}\left( 2,2\right) $%
-invariant polynomials ($Tr\mathbf{G}$\textbf{, }$Tr\left( \mathbf{G}%
^{2}\right) $, $Tr\left( \mathbf{G}^{3}\right) $, $Tr\left( \mathbf{G}%
^{4}\right) $) in the eigenvalues $\lambda _{i}$'s (or,
equivalently, in the charges $\mathcal{Q}_{1}$ and
$\mathcal{Q}_{2}$). As proved in App. \ref{Proof-Completeness}, such
four polynomials form a complete basis for the $SO_{h}^{v}\left( 2,2\right) $%
-invariant polynomials of the symmetric matrix $\mathbf{G}$.

\begin{enumerate}
\item  In the $stu$ model (and in the $\mathcal{N}=2$ Jordan symmetric
sequence for $n\geqslant 3$, as well as in the whole $\mathcal{N}=4$
infinite sequence (\ref{N=4}) for $n\geqslant 0$) there are no relations
among the four eigenvalues $\lambda _{i}$'s (generally, $\mathbf{G}$ has
rank $4$).

\item  In the $st^{2}$ model, since the charges can be arranged as an $%
SL_{v}\left( 2,\mathbb{R}\right) $-doublet of $SO\left( 2,1\right) $%
-vectors, the $4\times 4$ matrix $\mathbf{G}$ has non-maximal rank $3$, and
thus its determinant vanishes. From Eq. (\ref{d}), this yields the following
degree-$8$ constraint:
\begin{equation}
det\mathbf{G}\equiv d=0\Leftrightarrow \frac{1}{6}\left( Tr\mathbf{G}\right)
^{4}+\frac{1}{2}\left[ Tr\left( \mathbf{G}^{2}\right) \right] ^{2}+\frac{4}{3%
}Tr\left( \mathbf{G}^{3}\right) Tr\mathbf{G}-Tr\left( \mathbf{G}^{4}\right)
-Tr\left( \mathbf{G}^{2}\right) \left( Tr\mathbf{G}\right) ^{2}=0,
\label{detG=0}
\end{equation}
implying the vanishing of one eigenvalue of $\mathbf{G}$ itself. Eq. (\ref
{detG=0}) is an equivalent, manifestly $SO_{h}^{v}\left( 2,2\right) $%
-invariant re-writing of the degree-$8$ polynomial constraint (\ref{P8=0-2}%
), which can be used to eliminate \textit{e.g.} the highest-degree $\left[
SL_{h}\left( 2,\mathbb{R}\right) \times G_{4}\right] $-invariant of the set (%
\ref{4-2}), namely the $8$-degree Tr$\left( \frak{I}^{2}\right) $, in terms
of the other lower-degree invariants, thus reducing (\ref{4-2}) to the
following set of three:
\begin{equation}
\underset{\deg =2}{\mathcal{W}},~\underset{\deg =4}{\mathcal{X}},~\underset{%
\deg =6}{\mathbf{I}_{6}}.  \label{4-3}
\end{equation}
By using results (\ref{TrG})-(\ref{Tr(G^4)}) and Eqs. (\ref{W-call})-(\ref
{x-Frak}), the \textit{cubic} characteristic equation of $\mathbf{G}$ for $%
st^{2}$ model can be computed to read
\begin{equation}
det\left( \mathbf{G}-\lambda \mathbb{I}\right)
=\prod_{i=1}^{3}\left( \lambda -\lambda _{i}\right) =\lambda
^{3}+\mathcal{W}\lambda ^{2}+\left\| \mathbb{T}\right\| ^{2}\lambda
-\frac{1}{2}\mathbf{I}_{6}=0. \label{char-Eq-cubic}
\end{equation}

\item  In the $t^{3}$ model it further holds that (recall Eq. (\ref{t^3-rell}%
))
\begin{equation}
\mathcal{X}=0\Leftrightarrow \left\| \mathbb{T}\right\| ^{2}=\frac{1}{4}%
\mathcal{W}^{2}.  \label{X=0-2}
\end{equation}
Consequently, the set (\ref{4-3}) further reduces down to
\begin{equation}
\underset{\deg =2}{\mathcal{W}},~\underset{\deg =6}{\mathbf{I}_{6}},
\label{4-4}
\end{equation}
and the \textit{cubic} characteristic equation of $\mathbf{G}$ for $t^{3}$
model can be obtained from (\ref{char-Eq-cubic}) by implementing the further
condition (\ref{X=0-2}):
\begin{equation}
det\left( \mathbf{G}-\lambda \mathbb{I}\right) =\prod_{i=1}^{3}\left(
\lambda -\lambda _{i}\right) =\lambda \left( \lambda +\frac{1}{2}\mathcal{W}%
\right) ^{2}-\frac{1}{2}\mathbf{I}_{6}=0.  \label{char-Eq-cubic-t^3}
\end{equation}
Notice that in this model the rank of $\mathbf{G}$\ is still $3$.\medskip
\end{enumerate}

As mentioned above, the constraint relating the $\left[ SL_{h}\left( 2,%
\mathbb{R}\right) \times G_{4}\right] $-invariants Tr$\left( \frak{I}%
^{2}\right) $ and Tr$\left( \frak{I}^{3}\right) $ to the lower-degree
invariant polynomials are given by Eqs. (\ref{P12=0-2}) and (\ref{P8=0-2}).
Note that, as also discussed in Sec. \ref{Gen-Analysis-t^3}, by eliminating $%
\mathbf{I}_{6}$ in terms of Tr$\left( \frak{I}^{3}\right) $ increases the
degree in charges of the resulting polynomial constraints, from the degree-$%
12$ of (\ref{P12=0-2}) to the degree-$16$ of (\ref{P16=0-2}) and (\ref
{P16-t^3}).

\section{\label{Gen-Reducible-N=2-N=4}Generalization to $\mathcal{N}=2$
Jordan Symmetric Sequence\newline and $\mathcal{N}=4$ Theory}

Two infinite sequences of $d=4$ supergravity theories exhibit a factorised $%
U $-duality group and symmetric (vector multiplets') scalar
manifold, namely the Jordan symmetric sequence
(\ref{N=2-J-symm-seq}) and the $\mathcal{N}=4$ (generally
matter-coupled \cite{Bergshoeff, De Roo}) theory $\left( n\in
\mathbb{N}\cup \left\{ 0\right\}\right)$
\begin{equation}
\mathcal{N}=4:\frac{SL\left( 2,\mathbb{R}\right) }{U\left( 1\right)
}\times
\frac{SO\left( 6,n\right) }{SO\left( 6\right) \times SO\left( n\right) }%
~\left( \mathbb{R}\oplus \mathbf{\Gamma }_{5,n-1}\right) ,
\label{N=4}
\end{equation}
where (\ref{N=2-J-symm-seq}) is usually referred to as Jordan
symmetric sequence \cite{GST1,GST2}, and the round brackets in the
right-hand sides denote the corresponding \textit{reducible}
degree-$3$ Euclidean Jordan algebras \cite{GST1,GST2}. The number
$n_{V}$ of matter (vector) multiplets is given by $n+1$ in
(\ref{N=2-J-symm-seq}) and by $n$ in (\ref{N=4}).

The $stu$ and $st^{2}$ models respectively are the second ($n=2$)
and the first ($n=1$) elements of the sequence
(\ref{N=2-J-symm-seq}).

The result $I_{p=2}=7$ obtained for the $stu$ model in Sec. \ref
{Gen-Analysis-stu} can be proved to hold for the $\mathcal{N}=2$
Jordan
symmetric sequence (\ref{N=2-J-symm-seq}) with $n\geqslant 2$, and for the $%
\mathcal{N}=4$ theory coupled to any number $n\in \mathbb{N}$ of
matter (vector) multiplets.

Indeed, the $p=2$-center \textit{``large''} orbits with $\mathcal{I}%
_{4}\left( \mathcal{Q}_{1}\right) >0$, $\mathcal{I}_{4}\left( \mathcal{Q}%
_{2}\right) >0$ for the sequences (\ref{N=2-J-symm-seq}) and
(\ref{N=4}) respectively read:
\begin{eqnarray}
\mathbb{R}\oplus \mathbf{\Gamma }_{1,n-1} &:&\left\{
\begin{array}{l}
n=0,1,2:SL\left( 2,\mathbb{R}\right) \times SO\left( 2,n\right) ,~\text{BPS~(%
}n=0,1,2\text{),~non-BPS~(}n=1,2\text{)}; \\
\\
n\geqslant 3:SL\left( 2,\mathbb{R}\right) \times \frac{SO\left( 2,n\right) }{%
SO\left( n-2\right) },~\text{BPS,~non-BPS}; \\
\\
n\geqslant 4:SL\left( 2,\mathbb{R}\right) \times \frac{SO\left( 2,n\right) }{%
SO\left( 1,n-3\right) },~\text{non-BPS}; \\
\\
n\geqslant 5:SL\left( 2,\mathbb{R}\right) \times \frac{SO\left( 2,n\right) }{%
SO\left( 2,n-4\right) },~\text{non-BPS};
\end{array}
\right.   \label{p=2-1} \\
&&  \notag \\
\mathbb{R}\oplus \mathbf{\Gamma }_{5,n-1} &:&\left\{
\begin{array}{l}
n\geqslant 0:SL\left( 2,\mathbb{R}\right) \times \frac{SO\left( 6,n\right) }{%
SO\left( 2,n\right) },~\frac{1}{4}\text{-BPS}; \\
\\
n\geqslant 1:SL\left( 2,\mathbb{R}\right) \times \frac{SO\left( 6,n\right) }{%
SO\left( 3,n-1\right) },~\frac{1}{4}\text{-BPS}; \\
\\
n\geqslant 2:SL\left( 2,\mathbb{R}\right) \times \frac{SO\left( 6,n\right) }{%
SO\left( 4,n-2\right) },~\frac{1}{4}\text{-BPS,~non-BPS;} \\
\\
n\geqslant 3:SL\left( 2,\mathbb{R}\right) \times \frac{SO\left( 6,n\right) }{%
SO\left( 5,n-3\right) },~\text{non-BPS;} \\
\\
n\geqslant 4:SL\left( 2,\mathbb{R}\right) \times \frac{SO\left( 6,n\right) }{%
SO\left( 6,n-4\right) },~\text{non-BPS;}
\end{array}
\right.   \label{p=2-2}
\end{eqnarray}
By comparing these results with their $p=1$ counterparts \cite
{ADFT-rev,Kallosh-rev,BFGM2}, one can realize that the $2$-center
stabilizer is always contained into the $1$-center stabiliser with
corresponding supersymmetry-preserving properties. Furthermore, the
first line of (\ref {p=2-1}) summarizes the results of Sects.
\ref{Gen-Analysis-stu}-\ref {Gen-Analysis-t^3} on the $stu$ model
($n=2$) and its rank-$2$ and rank-$3$
descendants, namely the $st^{2}$ model ($n=1$) and the $t^{3}$ model ($n=0$%
). Since this latter does not belong to the Jordan symmetric
sequence, but
it is rather an isolated case in the classification of symmetric special K%
\"{a}hler manifolds (see \textit{e.g.} \cite{dWVVP} and Refs.
therein), the notation for $n=0$ in the first line of (\ref{p=2-1})
is only of formal nature.

The various orbits of (\ref{p=2-1}) and (\ref{p=2-2}) can be related
to the
possible choices of signs of the four eigenvalues $\lambda _{1}$,...,$%
\lambda _{4}$ of the Gramian matrix $\mathbf{G}$\textbf{\
}introduced in Sect. \ref{SO_h(2,2)}.

Moreover, (\ref{p=2-1}) and (\ref{p=2-2}) yield that the
$\mathcal{N}=2$ BPS orbit with $n=6$ matches the $\mathcal{N}=4$
non-BPS orbit with $n=2$, as
well as the $\mathcal{N}=2$ non-BPS orbit with $n=6$ matches the $\mathcal{N}%
=4$ $\frac{1}{4}$-BPS orbit with $n=2$. This can be traced back to
the fact that the corresponding theories share the very same bosonic
sector \cite {Gnecchi-1}.

Furthermore, it is worth remarking that the $n=0$ case of the
$\mathcal{N}=2$
Jordan symmetric sequence (\ref{N=2-J-symm-seq}) is nothing but the $%
\mathcal{N}=2$ \textit{axion-dilaton} model (truncation of the \textit{%
``pure''} $\mathcal{N}=4$ supergravity theory \cite{CSF}), whose
$2$-center split flow and marginal stability have been recently
studied (in a manifestly $U\left( 1,1\right) $-covariant symplectic
frame) in \cite {MS-FMO-1}.

Remarkably, the general formul\ae\
(\ref{general-1})-(\ref{charge-orbit-p}) yield that the number of
independent $G_{4}$-invariant polynomials in both
\textit{reducible} sequences (\ref{N=2-J-symm-seq}) and (\ref{N=4}) is $n$%
-independent and it amounts to $I_{p=2}=7$ :
\begin{eqnarray}
\mathbb{R}\oplus \mathbf{\Gamma }_{1,n-1}~\left( n\geqslant 2\right)
&:&I_{p=2}=4\left( 2+n\right) -3-\frac{\left( 2+n\right) \left( 1+n\right) }{%
2}+\frac{\left( n-2\right) \left( n-3\right) }{2}=7; \\
\mathbb{R}\oplus \mathbf{\Gamma }_{5,n-1}\left( n\in \mathbb{N}\cup
\left\{ 0\right\}\right)
&:&I_{p=2}=4\left( 6+n\right) -3-\frac{\left( 6+n\right) \left( 5+n\right) }{%
2}+\frac{\left( n+2\right) \left( n+1\right) }{2}=7.
\end{eqnarray}

Note that the symmetry (\ref{formula-1}) extends to
$SO_{h}^{v}\left( 2,2\right) \times SO\left( 2,n\right) $ and
$SO_{h}^{v}\left( 2,2\right)
\times SO\left( 6,n\right) $ for the sequences (\ref{N=2-J-symm-seq}) and (%
\ref{N=4}), respectively.\smallskip\

In a forthcoming investigation \cite{irreducible-1}, the analysis of $2$%
-center orbits and polynomial $G_{4}$-invariants will be extended to the $%
d=4 $ supergravity theories based on \textit{irreducible} rank-$3$
Euclidean
Jordan algebras, namely to $\mathcal{N}=2$ ``magic'' models and to $\mathcal{%
N}=5$, $6$, $8$ supergravities.

\section*{Acknowledgments}

We would like to thank Prof. V. S. Varadarajan and Prof. M. Trigiante for
enlightening discussions.

E. O. would like to thank CERN Theory Division for hospitality.

The work of S. F. is supported by the ERC Advanced Grant no. 226455, \textit{%
``Supersymmetry, Quantum Gravity and Gauge Fields''} (\textit{SUPERFIELDS})
and in part by DOE Grant DE-FG03-91ER40662.

The work of E. O. and A. Y. is supported by the ERC Advanced Grant no.
226455, \textit{``Supersymmetry, Quantum Gravity and Gauge Fields''} (%
\textit{SUPERFIELDS}).

\appendix

\section{\label{App-stu}Derivation of the Hierarchy of Constraints\newline
for $stu$, $st^{2}$ and $t^{3}$ Models}

This Appendix details the derivation of the polynomial constraint (\ref
{P12=0-2}) (or, equivalently, (\ref{P12=0-2})) for the $stu$ model. As
mentioned above, the further reduction of the constraint (\ref{P12=0-2}) to
the $st^{2}$ and $t^{3}$ models give rise to an hierarchy of manifestly $%
SL_{h}\left( 2,\mathbb{R}\right) $-invariant polynomial relations among the
various $G_{4}$-invariant polynomials for these models.


We work in the ``special coordinates'' symplectic frame. In order to
highlight the actual dependence of the $G_{4}$-invariant polynomials on the
charges $\mathcal{Q}_{1}$ and $\mathcal{Q}_{2}$ themselves, we start and
define some charge variables invariant under the $SO\left( 1,1\right) $
rescaling symmetry of $G_{4}$ \cite{dWVVP}. Indeed, by recalling that the
components of $\mathcal{Q}_{1}$ and $\mathcal{Q}_{2}$ have the following $%
SO\left( 1,1\right) $-weights ($i=1,2,3$):
\begin{eqnarray}
&&
\begin{array}{c}
SO\left( 1,1\right) : \\
\mathcal{Q}_{1}=
\end{array}
\begin{array}{cccc}
+3 & +1 & -3 & -1 \\
(p^{0}, & p^{i}, & q_{0}, & q_{i});
\end{array}
\\
&&
\begin{array}{c}
SO\left( 1,1\right) : \\
\mathcal{Q}_{2}=
\end{array}
\begin{array}{cccc}
+3 & +1 & -3 & -1 \\
(P^{0}, & P^{i}, & Q_{0}, & Q_{i}),
\end{array}
\end{eqnarray}
one can define
\begin{eqnarray}
&&\mathbf{x} \equiv q_{1}p^{1},\;\mathbf{y}\equiv q_{2}p^{2},\;\mathbf{z}%
\equiv q_{3}p^{3},\;\mathbf{r}\equiv q_{0}p^{0},  \notag \\
&&\mathbf{X}\equiv Q_{1}P^{1},\;\mathbf{Y}\equiv Q_{2}P^{2},\;\mathbf{Z}%
\equiv Q_{3}P^{3},\;\mathbf{R}\equiv Q_{0}P^{0},  \notag \\
&&\mathbf{Z}_{1}\equiv Q_{0}P^{1}P^{2}P^{3},\quad \mathbf{z}_{1}\equiv
q_{0}p^{1}p^{2}p^{3},  \notag \\
&&\mathbf{u}\equiv \frac{q_{0}}{Q_{0}},\;\mathbf{u}_{2}\equiv \frac{p^{2}}{%
P^{2}},\;\mathbf{u}_{3}\equiv \frac{p^{3}}{P^{3}}.
\label{SO(1,1)-inv-charges}
\end{eqnarray}
It can be explicitly shown that the eight polynomial $\left[ SL\left( 2,%
\mathbb{R}\right) \right] ^{3}$-invariants $\mathbf{I}_{+2}$, $\mathbf{I}%
_{+1}$, $\mathbf{I}^{\prime }$, $\mathbf{I}^{\prime \prime }$, $\mathbf{I}%
_{-1}$, $\mathbf{I}_{-2}$, $I_{6}$ and $\mathcal{W}$ of the $stu$ model can
be re-written only in terms of the thirteen $SO\left( 1,1\right) $-invariant
charge variables defined by (\ref{SO(1,1)-inv-charges}).

As stated at the start of\ Sec. \ref{stu-Descendants}, all minors of order $%
8 $ of the relevant Jacobian matrix $\mathbf{J}$ defined by (\ref{Jac}), (%
\ref{I-stu}) and (\ref{Qbold-alpha-stu}) do vanish. On the other hand, all
minors of order $7$ are non-zero. This is a compelling evidence that the
number of \textit{independent} $\left[ SL\left( 2,\mathbb{R}\right) \right]
^{3}$-invariant polynomials is seven.

Thus, a (polynomial) relation constraining the aforementioned eight
invariants is expected to exist. In order to derive it, it is convenient to
work in a particularly simple charge configuration, in which some of the $%
SO\left( 1,1\right) $-invariant charge variables (\ref{SO(1,1)-inv-charges})
vanish; namely:
\begin{equation}
p^{1}=q_{1}=p^{2}=q_{2}=p^{3}=q_{3}=0\Rightarrow \mathbf{x}=\mathbf{z}=%
\mathbf{y}=\mathbf{z}_{1}=\mathbf{u}_{2}=\mathbf{u}_{3}=\frac{\mathbf{x\,y\,z%
}}{\mathbf{z}_{1}}=\frac{\mathbf{z}_{1}}{\mathbf{u}_{2}}=\frac{\mathbf{z}_{1}%
}{\mathbf{u}_{3}}=0.  \label{choice-1}
\end{equation}
This implies the extremal BH at center $1$ to be non-BPS $Z_{H}\neq 0$ (%
\textit{i.e.} $\mathbf{I}_{+2}<0$), thus resulting in
\begin{equation}
\mathbf{r}=\sqrt{-\mathbf{I}_{+2}}.  \label{choice-2}
\end{equation}
Clearly, these choices imply some loss in generality, but remarkably, as we
will see below, the results achieved within such a configuration actually
hold for a completely generic $2$-center charge configuration.

By plugging (\ref{choice-1}) and (\ref{choice-2}) into the expressions of $%
\mathbf{I}_{+2}$, $\mathbf{I}_{+1}$, $\mathbf{I}^{\prime }$, $\mathbf{I}%
^{\prime \prime }$, $\mathbf{I}_{-1}$, $\mathbf{I}_{-2}$, $I_{6}$ and $%
\mathcal{W}$ of the $stu$ model, one can solve the resulting algebraic Eqs.
for $\mathbf{X}$, $\mathbf{Y}$, $\mathbf{u}$, $\mathbf{R}$ and $\mathbf{Z}%
_{1}$ as follows:
\begin{eqnarray}
\mathbf{u} &=&\frac{2\mathbf{I}_{+2}}{2\mathbf{I}_{+1}+\mathcal{W}\sqrt{-%
\mathbf{I}_{+2}}},\quad \mathbf{R}=\frac{4\mathbf{I}_{+1}^{2}+\mathbf{I}_{+2}%
\mathcal{W}^{2}}{4(-\mathbf{I}_{+2})^{3/2}},  \notag \\
\mathbf{X} &=&-\frac{4\mathbf{I}_{+1}^{2}+\mathbf{I}_{+2}\left( \mathcal{W}%
^{2}-4\mathbf{I}^{\prime }\right) }{4(-\mathbf{I}_{+2})^{3/2}},\quad \mathbf{%
Y}=2\frac{\mathbf{I}_{+2}\mathbf{I}^{\prime \prime }-\mathbf{I}_{+1}^{2}}{(-%
\mathbf{I}_{+2})^{3/2}}-\mathbf{Z},  \label{sol} \\
\mathbf{Z}_{1} &=&\frac{1}{8\mathbf{I}_{+2}^{3}}\left\{ 8\mathbf{I}_{+1}^{4}+%
\mathbf{I}_{+2}^{2}\left[ 2\mathbf{I}_{-1}\mathbf{I}_{+1}+\mathcal{W}\left(
\mathbf{I}_{6}+\mathbf{I}_{-1}\sqrt{-\mathbf{I}_{+2}}-\mathbf{I}^{\prime
\prime }\mathcal{W}\right) \right] \right.  \notag \\
&&\qquad \quad \left. -\mathbf{I}_{+2}\mathbf{I}_{-1}\left[ 2\mathbf{I}_{6}%
\sqrt{-\mathbf{I}_{+2}}+\mathbf{I}^{\prime }\mathcal{W}\sqrt{-\mathbf{I}_{+2}%
}+\mathbf{I}_{+1}\left( 2\mathbf{I}^{\prime }+4\mathbf{I}^{\prime \prime }-%
\mathcal{W}^{2}\right) \right] \right\} .  \label{sol-2}
\end{eqnarray}
By so doing, one ends up with following equation for the remaining charge
variable $\mathbf{Z}:$
\begin{equation}
\mathbf{Z}^{2}=\frac{1}{4(\mathbf{I}_{+2})^{3}}\left\{ 4\mathbf{I}_{+1}^{4}-%
\mathbf{I}_{-2}\mathbf{I}_{+2}^{3}-8\mathbf{I}_{+2}\mathbf{I}^{\prime \prime
}\mathbf{I}_{+1}^{2}+\mathbf{I}_{+2}^{2}\left[ (\mathbf{I}^{\prime }-2%
\mathbf{I}^{\prime \prime })^{2}+4\mathbf{I}_{+1}\mathbf{I}_{-1}+2\mathbf{I}%
_{6}\mathcal{W}\right] \right\} +2\frac{\left( \mathbf{I}_{+2}\mathbf{I}%
^{\prime \prime }-\mathbf{I}_{+1}^{2}\right) \mathbf{Z}}{(-\mathbf{I}%
_{+2})^{3/2}},  \label{Z^2}
\end{equation}
and self-consistency condition given by Eq. (\ref{P12=0-2}).

Remarkably, when relaxing the conditions (\ref{choice-1})-(\ref{choice-2}),
one can check by direct calculation that Eq. (\ref{P12=0-2}) holds for a
completely general $2$-center charge configuration.

In order to reduce the $stu$ model to the $st^{2}$ model, the following
identifications of charges are to be performed (within the positions (%
\ref{choice-1})-(\ref{choice-2})):
\begin{equation}
P^{3}=P^{2}\equiv P^{2},\quad Q_{3}=Q_{2}\equiv \frac{Q_{2}}{2}.
\end{equation}
This implies $\mathbf{Y}=\mathbf{Z}$, where (\ref{sol}) implies
\begin{equation}
\mathbf{Y}=\mathbf{Z}=\frac{\mathbf{I}_{+2}\mathbf{I}^{\prime \prime }-%
\mathbf{I}_{+1}^{2}}{(-\mathbf{I}_{+2})^{3/2}}.  \label{Y=Z}
\end{equation}
By inserting (\ref{Y=Z}) into (\ref{Z^2}), the result (\ref{P8=0-2}) is
achieved. By plugging this latter back into (\ref{P12=0-2}), the $16$-degree
constraint (\ref{P16=0-2})-(\ref{P16-t^3}) is obtained.

On the other hand, the reduction of the $stu$ model down to the $t^{3}$
model entails the following charge identifications:
\begin{equation}
P^{3}=P^{2}=P^{1}\equiv P,\quad Q_{3}=Q_{2}=Q_{1}\equiv \frac{Q}{3}.
\end{equation}
These latter imply $\mathbf{X}=\mathbf{Y}=\mathbf{Z}$, and Eqs. (\ref{sol})-(%
\ref{sol-2}) thus yield Eq. (\ref{t^3-rell}). This latter, inserted into
Eqs. (\ref{P8=0-2}) and (\ref{P16=0-2}), respectively gives (\ref{P8=0-3})
and (\ref{P16-t^3=0}).

Once again, all above results can be checked to hold in a general $2$-charge
configuration, and therefore they are completely general.\smallskip

This analysis relates a ``minimal'' set of BH charges to the ``minimal''
number of independent $G_{4}$-invariant polynomials discussed in the present
paper (see in particular Sec. \ref{stu-Descendants}).

Finally, the action of the $d=4$ $U$-duality group $G_{4}$ on the
charge vectors $\mathcal{Q}_{1}$ and $\mathcal{Q}_{2}$ can be
summarised as follows:

\begin{enumerate}
\item  In the rank-$3$ $stu$ model ($G_{4}=\left[ SL\left( 2,\mathbb{R}%
\right) \right] ^{3}$) there are $7$ independent polynomial $G_{4}$%
-invariants depending on the $13$ $SO\left( 1,1\right) $-invariant
combinations. Thus, out of the $16$ charges composing the charge vector $%
\mathbf{Q}^{\alpha }$ defined in (\ref{Qbold-alpha-stu}), $6$ charges can be
set to zero by the action of the non-dilatational $6$ generators of $G_{4}$,
spanning $\left[ \frac{SL\left( 2,\mathbb{R}\right) }{SO\left( 1,1\right) }%
\right] ^{3}$ (namely, one conformal boost and one translational generator
for each of the three $\frac{SL\left( 2,\mathbb{R}\right) }{SO\left(
1,1\right) }$ factors). By recalling the definitions (\ref
{SO(1,1)-inv-charges}) of the $SO\left( 1,1\right) $-invariant charge
variables, a representative of a ``minimal'' charge configuration is \textit{%
e.g.} given by
\begin{eqnarray}
\mathcal{Q}_{1} &=&\left( \frac{\mathbf{r}%
}{\mathbf{u Z_{1}}}P^{1}P^{2}P^{3},\frac{\mathbf{z}_{1}}{\mathbf{Z}%
_{1}\mathbf{u u}_{2}\mathbf{u}_{3}}P^{1},\mathbf{u_{2}}P^{2},\mathbf{u_{3}}P^{3},\frac{\mathbf{u{Z}_{1}}}{P^{1}P^{2}P^{3}},%
\frac{ \mathbf{x Z}_{1}}{\mathbf{z}_{1}}\frac{\mathbf{u u}_{2}\mathbf{u}_{3}}{P^1},%
0,0\right) ^{T};  \notag \\
\mathcal{Q}_{2} &=&\left( 0,P^{1},P^{2},P^{3},\frac{\mathbf{Z}_{1}}{%
P^{1}P^{2}P^{3}},0,0,0\right) ^{T},  \label{stu-minimal}
\end{eqnarray}
with $\mathbf{Z}_{1}>0\Rightarrow \mathcal{I}_{4}\left(
\mathcal{Q}_{2}\right) >0$, and $\mathcal{I}_{4}\left( \mathcal{Q}%
_{1}\right) >0$. In (\ref{stu-minimal}) $P^{1},P^{2},P^{3}\in \mathbb{R}%
_{0}^{+}$ are the parameters of the three $SO(1,1)$ (generated by the three
non-compact Cartan generators of $G_{4}=\left[ SL\left( 2,\mathbb{R}\right) %
\right] ^{3}$).

\item  In the rank-$2$ $st^{2}$ model ($G_{4}=\left[ SL\left( 2,\mathbb{R}%
\right) \right] ^{2}$) there are $6$ independent polynomial $G_{4}$%
-invariants depending on the $10$ $SO\left( 1,1\right) $-invariant
combinations. Thus, out of the $12$ charges composing the charge vector $%
\mathbf{Q}^{\alpha }$ defined in (\ref{Qbold-alpha-st^2}), $4$ charges can
be set to zero by the action of the non-dilatational $4$ generators of $%
G_{4} $, spanning $\left[ \frac{SL\left( 2,\mathbb{R}\right) }{SO\left(
1,1\right) }\right] ^{2}$ (namely, one conformal boost and one translational
generator for each of the two $\frac{SL\left( 2,\mathbb{R}\right) }{SO\left(
1,1\right) }$ factors). By recalling the definitions (\ref
{SO(1,1)-inv-charges}) of the $SO\left( 1,1\right) $-invariant charge
variables, a representative of a ``minimal'' charge configuration is \textit{%
e.g.} given by
\begin{eqnarray}
\mathcal{Q}_{1} &=&\left( \frac{\mathbf{r}%
}{\mathbf{u Z_{1}}}P^{1}\left(
P^{2}\right) ^{2},\frac{\mathbf{z}_{1}}{\mathbf{Z}%
_{1}\mathbf{u u_{2}^2}}P^{1},\mathbf{u_{2}}P^{2},\frac{\mathbf{u{Z}_{1}}}{P^{1}\left(
P^{2}\right) ^{2}},%
\frac{ \mathbf{x Z}_{1}}{\mathbf{z}_{1}}\frac{\mathbf{u u_{2}^2}}{P^1},%
0\right) ^{T};  \notag \\
\mathcal{Q}_{2} &=&\left( 0,P^{1},P^{2},\frac{\mathbf{Z}_{1}}{P^{1}\left(
P^{2}\right) ^{2}},0,0\right) ^{T},  \label{st^2-minimal}
\end{eqnarray}
with $\mathbf{Z}_{1}>0\Rightarrow \mathcal{I}%
_{4}\left( \mathcal{Q}_{2}\right) >0$, and $\mathcal{I}_{4}\left( \mathcal{Q}%
_{1}\right) >0$. In (\ref{st^2-minimal}) $P^{1},P^{2}\in \mathbb{R}_{0}^{+}$
are the parameters of the two $SO(1,1)$ (generated by the two non-compact
Cartan generators of $G_{4}=\left[ SL\left( 2,\mathbb{R}\right) \right] ^{2}$%
).

\item  In the rank-$1$ irreducible $t^{3}$ model ($G_{4}=SL\left( 2,\mathbb{R%
}\right) $) there are $5$ independent polynomial $G_{4}$-invariants
depending on the $7$ $SO\left( 1,1\right) $-invariant combinations. Thus,
out of the $8$ charges composing the charge vector $\mathbf{Q}^{\alpha }$
defined in (\ref{Q-bold-alpha-t^3}), $2$ charges can be set to zero by the
action of the non-dilatational $2$ generators of $G_{4}$, spanning $\frac{%
SL\left( 2,\mathbb{R}\right) }{SO\left( 1,1\right) }$ (namely, one conformal
boost and one translational generator). By recalling the definitions (\ref
{SO(1,1)-inv-charges}) of the $SO\left( 1,1\right) $-invariant charge
variables, a representative of a ``minimal'' charge configuration is \textit{%
e.g.} given by
\begin{eqnarray}
\mathcal{Q}_{1} &=&\left( \frac{\mathbf{r}%
}{\mathbf{u Z_{1}}}\left(
P\right) ^{3},\mathbf{u_{2}}P,\frac{\mathbf{u{Z}_{1}}}{\left(
P\right) ^{3}},%
3\frac{ \mathbf{x}}{\mathbf{u_{2}}P}\right) ^{T};  \notag \\
\mathcal{Q}_{2} &=&\left( 0,P,\frac{\mathbf{Z}_{1}}{\left(
P\right) ^{3}},0\right) ^{T},  \label{t^3-minimal}
\end{eqnarray}
with $\mathbf{Z}_{1}>0\Rightarrow \mathcal{I}_{4}\left( \mathcal{Q}%
_{2}\right) >0$, and $\mathcal{I}_{4}\left( \mathcal{Q}_{1}\right) >0$. In (%
\ref{t^3-minimal}) $P\in \mathbb{R}_{0}^{+}$ is the parameter of the $SO(1,1)
$ generated by the non-compact Cartan generator of $G_{4}=SL\left( 2,\mathbb{%
R}\right) $.
\end{enumerate}

\section{\label{App-t^3}Constrained Calabi-Vesentini Basis for $t^{3}$ Model}

As mentioned in Sec. \ref{Gen-Analysis-t^3}, the application of the
Calabi-Vesentini \cite{CV-original,CDFVP-1} $\mathbb{T}$-tensor
formalism to the $t^{3}$ model deserves a separate treatment, which
we are going detail in the present Appendix.

In order to deal with this, let us start and recall some basic facts
on the Calabi-Vesentini (CV) basis of $\mathcal{N}=2$, $d=4$ Jordan
symmetric sequence. In particular, let us consider the $stu$ model,
whose CV basis has been explicitly discussed also in \cite{BKRSW}.
Thus, for such a model, the symplectic sections are manifestly
covariant under the whole $U$-duality group $SL\left(
2,\mathbb{R}\right) \times SO\left( 2,2\right) \sim \left[ SL\left(
2,\mathbb{R}\right) \right] ^{3}$ ($\Lambda =0,1,2,3$)
\begin{equation}
\mathbf{V}\left( s,u_{1},u_{2}\right) \mathbf{\equiv }\left(
\begin{array}{c}
X^{\Lambda } \\
\\
F_{\Lambda }
\end{array}
\right) \equiv \left(
\begin{array}{c}
X^{\Lambda }(u_{1},u_{2}) \\
\\
s\eta _{\Lambda \Sigma }X^{\Sigma }(u_{1},u_{2})
\end{array}
\right) ,  \label{Omega}
\end{equation}
where $X^{\Lambda }(u)$ satisfies the condition $X^{\Lambda }(u)\eta
_{\Lambda \Sigma }X^{\Sigma }(u)=0$. The axion-dilaton field $s$
parameterizes the coset $\frac{SL(2,\mathbb{R})}{U(1)}$, whereas the two
independent complex coordinates $u_{1},u_{2}$ parameterize the coset $\frac{%
SO(2,2)}{SO(2)\times SO(2)}$. Note that, as shown in \cite{CDFVP-1}, in this
symplectic frame a prepotential does not exist at all; however, it is still
possible to calculate all the relevant geometrical quantities, using the
standard formul\ae\ of special K\"{a}hler geometry \cite{CDFVP-1}.

The relation between the CV symplectic frame specified by Eq. (\ref{Omega})
and the ``special coordinates'' symplectic frame (whose manifest covariance
is restricted to the $d=5$ $U$-duality group $\left[ SO(1,1)\right] ^{2}$;
see \textit{e.g.} the treatment in \cite{BKRSW}) is given by \cite
{CDFVP-1,BKRSW} (``SC'' is acronym for ``special coordinates'')
\begin{equation}
X_{CV}^{\Lambda }=\frac{1}{\sqrt{2}}\left(
\begin{array}{c}
1-t\,u \\
-(t+u) \\
-(1+t\,u) \\
t-u
\end{array}
\right) .  \label{X-het-stu}
\end{equation}
Correspondingly, the BH charges in both symplectic frames are related by the
following $Sp\left( 8,\mathbb{R}\right) $ finite transformation \cite
{CDFVP-1,BKRSW}
\begin{equation}
\left(
\begin{array}{c}
p^{\Lambda } \\
\\
q_{\Lambda }
\end{array}
\right) _{stu,CV}=\frac{1}{\sqrt{2}}\left(
\begin{array}{c}
p^{0}-q_{1} \\
-p^{2}-p^{3} \\
-p^{0}-q_{1} \\
p^{2}-p^{3} \\
p^{1}+q_{0} \\
-q_{2}-q_{3} \\
p^{1}-q_{0} \\
q_{2}-q_{3}
\end{array}
\right) _{stu,SC}.  \label{charge STU}
\end{equation}

Starting from the CV basis of the $stu$ model introduced above, the
manifestly $\left( SL\left( 2,\mathbb{R}\right) \times SO\left( 2,1\right)
\right) $-covariant CV basis for the $st^{2}$ model can be obtained by
performing the following charge identifications\footnote{%
For a discussion of the \textit{``}$stu\rightarrow st^{2}\rightarrow t^{3}$%
\textit{\ degeneration''} in a different symplectic frame (relevant for
applications to Quantum Information Theory \cite{QIT-Refs}), see \textit{e.g.%
} the discussion in Sec. 5 of \cite{BMOS-1}.}:
\begin{equation}
\left\{
\begin{array}{l}
p_{stu,SC}^{2}=p_{stu,SC}^{3}\equiv p_{st^{2},SC}^{2}; \\
\\
q_{2,stu,SC}=q_{3,stu,SC}\equiv \frac{1}{2}q_{2,stu,SC}.
\end{array}
\right.  \label{charge-ids}
\end{equation}
By so doing, the $Sp\left( 8,\mathbb{R}\right) $ transformation (\ref{charge
STU}) reduces to the following $Sp\left( 6,\mathbb{R}\right) $
transformation ($\Lambda =0,1,2$):
\begin{equation}
\left(
\begin{array}{c}
p^{\Lambda } \\
\\
q_{\Lambda }
\end{array}
\right) _{st^{2},CV}=\frac{1}{\sqrt{2}}\left(
\begin{array}{c}
p^{0}-q_{1} \\
-2p^{2} \\
-p^{0}-q_{1} \\
p^{1}+q_{0} \\
-q_{2} \\
p^{1}-q_{0}
\end{array}
\right) _{st^{2},SC}.  \label{chargeST^2}
\end{equation}

On the other hand, if one wants to adapt the CV basis for $stu$ model
introduced above to the $t^{3}$ model, the following charge identifications
are to be performed:
\begin{equation}
\left\{
\begin{array}{l}
p_{stu,SC}^{1}=p_{stu,SC}^{2}=p_{stu,SC}^{3}\equiv p_{t^{3},SC}^{1}; \\
\\
q_{1,stu,SC}=q_{2,stu,SC}=q_{3,stu,SC}\equiv \frac{1}{3}q_{1,t^{3},SC}.
\end{array}
\right.  \label{charge-ids-2}
\end{equation}
Through these identifications, the $Sp\left( 8,\mathbb{R}\right) $
transformation (\ref{charge STU}) reduces to the following $Sp\left( 6,%
\mathbb{R}\right) $ transformation ($\Lambda =0,1,2$)
\begin{equation}
\left(
\begin{array}{c}
p^{\Lambda } \\
\\
q_{\Lambda }
\end{array}
\right) _{t^{3},CV}=\frac{1}{\sqrt{2}}\left(
\begin{array}{c}
p^{0}-q_{1}/3 \\
-2p^{1} \\
-p^{0}-q_{1}/3 \\
p^{1}+q_{0} \\
-2q_{1}/3 \\
p^{1}-q_{0}
\end{array}
\right) _{t^{3},SC}.  \label{chargeT^3}
\end{equation}
Notice that still the symplectic index $\Lambda $ runs $0,1,2$, thus there
would be six charges, not consistent with the four magnetic and electric
charges of the $t^{3}$ model. In fact, the $p^{\Lambda }$ and $q_{\Lambda }$
$3$-vectors of (\ref{chargeT^3}) in the CV basis are not independent, but
rather they are constrained by the two relations
\begin{equation}
t^{3}\text{,~}CV:\left\{
\begin{array}{l}
p^{0}+p^{2}-q_{1}=0; \\
\\
q_{0}+q_{2}+p^{1}=0.
\end{array}
\right.  \label{t^3-constraints}
\end{equation}
This yields to a consistent counting, because two real $3$-vectors $%
p^{\Lambda }$ and $q_{\Lambda }$ with two real constraints (\ref
{t^3-constraints}) corresponds to four real charge degrees of freedom,
namely the four charges of the $t^{3}$ model itself. Eqs. (\ref{chargeT^3})
and (\ref{t^3-constraints}) defined a ``constrained'' CV symplectic frame
for the $t^{3}$ model.

Interestingly, the relations (\ref{t^3-constraints}) can be recast in a
covariant fashion using a real form of the Pauli matrices
\begin{equation}
\sigma _{1}\equiv \left(
\begin{array}{cc}
0 & 1 \\
1 & 0
\end{array}
\right) ,\quad \sigma _{2}\equiv \left(
\begin{array}{cc}
1 & 0 \\
0 & -1
\end{array}
\right) ,\quad \sigma _{3}\equiv \left(
\begin{array}{cc}
0 & -1 \\
1 & 0
\end{array}
\right) ,  \label{sigmas}
\end{equation}
and $3$-vector of $2$-component spinors
\begin{equation}
r^{\Lambda }=\left(
\begin{array}{c}
p^{\Lambda } \\
\eta ^{\Lambda \Sigma }q_{\Sigma }
\end{array}
\right) \equiv \left(
\begin{array}{c}
\overrightarrow{r}^{1} \\
\overrightarrow{r}^{2}
\end{array}
\right) ,  \label{spinors}
\end{equation}
where the definition of the $SO\left( 2,1\right) $-vectors $\overrightarrow{s%
}^{\alpha }$ ($\alpha =1,2$) is done for later convenience. By means
of (\ref {sigmas}) and (\ref{spinors}), the constraints
(\ref{t^3-constraints}) can be recast as\footnote{Note that the quartic invariant polynomial of the $\mathbf{4}$ (spin $s=3/2$%
) irrepr. of $SL\left( 2,\mathbb{R}\right) $ can be written as \cite
{Procesi-book}
\begin{equation*}
\mathcal{I}_{4}=\left| \overrightarrow{r}^{1}\times \overrightarrow{r}%
^{2}\right| ^{2}=\left| \overrightarrow{r}^{1}\right| ^{2}\left|
\overrightarrow{r}^{2}\right| ^{2}-\left(
\overrightarrow{r}^{1}\cdot \overrightarrow{r}^{2}\right) ^{2},
\end{equation*}
constrained by (\ref{t^3-constraints-2}), where ``$\times $''
denotes the
exterior product of the $3$-vectors $\overrightarrow{r}^{1}$ and $%
\overrightarrow{r}^{2}$, and the square norms are computed with the $%
SO\left( 2,1\right) $-metric $\eta ^{\Lambda \Sigma }$. In the
``special
coordinates'' symplectic frame used in Sect. \ref{Gen-Analysis-t^3}, $%
\mathcal{I}_{4}$ is given by Eq. (\ref{I4-t^3}).}
\begin{equation}
\vec{\sigma}\cdot \,\overrightarrow{r}=0.  \label{t^3-constraints-2}
\end{equation}

On a group theoretical perspective, the constraints (\ref{t^3-constraints})
or (\ref{t^3-constraints-2}) denote the projection on the $\mathbf{4}$ (spin
$s=3/2$) irrepr. in the tensor product of the irreprs. $\mathbf{3}$ (spin $%
s=1$) and $\mathbf{2}$ (spin $s=1/2$) of the $d=4$ $U$-duality group $%
SL\left( 2,\mathbb{R}\right) $ of the $t^{3}$ model:
\begin{equation}
\mathit{``}st^{2}\rightarrow t^{3}\mathit{\ }\text{\textit{reduction}}%
\mathit{":~}\mathbf{3}\times \mathbf{2}=\fbox{$\underset{t^{3}}{\mathbf{4}}$}%
+\fbox{$\underset{\text{projected out by }\vec{\sigma}\cdot \,%
\overrightarrow{r}=0}{\mathbf{2}}$}\mathbf{.}
\end{equation}
This is consistent with the above treatment, because the tensor product $%
\mathbf{3}\times \mathbf{2}$ realizes the \textit{``}$st^{2}\rightarrow
t^{3} $\textit{\ reduction''} of the charge repr. $\left( \mathbf{3},\mathbf{%
2}\right) $ of the $U$-duality $SL\left( 2,\mathbb{R}\right) \times SO\left(
2,1\right) \sim \left[ SL\left( 2,\mathbb{R}\right) \right] ^{2}$ of the $%
st^{2}$ model down to the charge irrepr. $\mathbf{4}$ of the $U$-duality $%
SL\left( 2,\mathbb{R}\right) $ of the $t^{3}$ model itself.\bigskip

In the case of an extremal BH $2$-center solution, within this
``constrained'' CV symplectic frame, one can consider the $\mathbb{T}$%
-tensor formalism introduced above for the $t^{3}$, by simply considering
the $st^{2}$ model in CV basis and implementing the constraints (\ref
{t^3-constraints}) (or, equivalently, (\ref{t^3-constraints-2})).

The center $1$ is constrained by (\ref{t^3-constraints-2}), whereas the
center $2$ is constrained by
\begin{eqnarray}
\vec{\sigma}\cdot \,\overrightarrow{s} &=&0;  \label{t^3-constraints-3} \\
s^{\Lambda } &=&\left(
\begin{array}{c}
P^{\Lambda } \\
\eta ^{\Lambda \Sigma }Q_{\Sigma }
\end{array}
\right) \equiv \left(
\begin{array}{c}
\overrightarrow{s}^{1} \\
\overrightarrow{s}^{2}
\end{array}
\right) .  \label{spinorss-2}
\end{eqnarray}
A consequence of relations (\ref{t^3-constraints-2}) and (\ref
{t^3-constraints-3}) can be proved to be
\begin{equation}
4\left( p\cdot P\right) \left( q\cdot Q\right) -4\left( p\cdot Q\right)
\left( P\cdot q\right) -\left( \overrightarrow{p}\times \overrightarrow{Q}-%
\overrightarrow{q}\times \overrightarrow{P}\right) ^{2}=\left( p\cdot
Q-P\cdot q\right) ^{2},  \label{resss}
\end{equation}
where ``$\times $'' denotes the exterior product of the $3$-vectors $%
\overrightarrow{p}\equiv p^{\Lambda }$, $\overrightarrow{Q}\equiv Q^{\Lambda
}$, \textit{etc.}, and the square in the last term in the l.h.s. is
performed with the $SO\left( 2,1\right) $-metric $\eta ^{\Lambda \Sigma }$.
Note that, by means of definitions (\ref{I'-CV}) and (\ref{I''-CV}), (\ref
{resss}) is equivalent to the vanishing of the $SL_{h}\left( 2,\mathbb{R}%
\right) $-singlet $\mathcal{X}$ (see Eq. (\ref{t^3-rell})).

In order to prove (\ref{resss}), we will work with complex groups. By
recalling that Greek lowercase indices are spinor (\textit{e.g.} $\alpha
=1,2 $) - whereas Latin lowercase indices are vector (\textit{e.g.} $i=1,2,3$%
) - of $SL\left( 2,\mathbb{C}\right) $, one starts and introduces the spinor
\begin{equation}
\left( \overrightarrow{\sigma }\cdot \overrightarrow{r}\right) _{\alpha
}\equiv \sigma _{~\alpha \beta }^{i}r_{i}^{\beta },
\end{equation}
and the vector-spinor:
\begin{equation}
\left( \overrightarrow{\sigma }~\overrightarrow{\sigma }\cdot
\overrightarrow{s}\right) _{\alpha }^{i}\equiv \sigma _{~\alpha \beta
}^{i}\sigma _{~\gamma }^{j\mid \beta }s_{j}^{\gamma }.
\end{equation}
By expanding the products of $\sigma $-matrices through the Fierz
identities, and elaborating the $\epsilon _{ijk}$-symbols, one obtains that
\begin{eqnarray}
\left( \overrightarrow{\sigma }\cdot \overrightarrow{r}\right) _{\alpha
}\left( \overrightarrow{\sigma }~\overrightarrow{\sigma }\cdot
\overrightarrow{s}\right) _{\beta }\epsilon ^{\alpha \beta } &=&-4i\left(
\overrightarrow{r}^{\alpha }\times \overrightarrow{s}^{\beta }\right)
\epsilon _{\alpha \beta }-4\left( \overrightarrow{r}^{\alpha }\cdot
\overrightarrow{s}^{\beta }\right) \overrightarrow{\sigma }_{\alpha \beta }
\notag \\
&&+4\left( \overrightarrow{\sigma }_{\alpha \beta }\cdot \overrightarrow{r}%
^{\alpha }\right) \overrightarrow{s}^{\beta }-2\left( \overrightarrow{\sigma
}_{\alpha \beta }\cdot \overrightarrow{s}^{\alpha }\right) \overrightarrow{r}%
^{\beta },
\end{eqnarray}
which, under the constraints (\ref{t^3-constraints-2}) and (\ref
{t^3-constraints-3}), yields
\begin{equation}
i\left( \overrightarrow{r}^{\alpha }\times \overrightarrow{s}^{\beta
}\right) \epsilon _{\alpha \beta }=-\left( \overrightarrow{r}^{\alpha }\cdot
\overrightarrow{s}^{\beta }\right) \overrightarrow{\sigma }_{\alpha \beta }.
\label{intermediate-1}
\end{equation}
By squaring (\ref{intermediate-1}) and using the Fierz identities, the
following result is finally achieved:
\begin{gather}
-\left( \overrightarrow{r}_{1}\times \overrightarrow{s}_{2}-\overrightarrow{r%
}_{2}\times \overrightarrow{s}_{1}\right) ^{2}=\left( \overrightarrow{r}%
^{\alpha }\cdot \overrightarrow{s}^{\beta }\right) \left( \overrightarrow{r}%
^{\gamma }\cdot \overrightarrow{s}^{\delta }\right) \overrightarrow{\sigma }%
_{\alpha \beta }\cdot \overrightarrow{\sigma }_{\gamma \delta }  \notag \\
=\left( \overrightarrow{r}_{1}\cdot \overrightarrow{s}_{2}-\overrightarrow{r}%
_{2}\cdot \overrightarrow{s}_{1}\right) ^{2}-4\left( \overrightarrow{r}%
_{1}\times \overrightarrow{r}_{2}\right) \cdot \left( \overrightarrow{s}%
_{1}\times \overrightarrow{s}_{2}\right) \\
\Updownarrow  \notag \\
4\left( \overrightarrow{r}_{1}\times \overrightarrow{r}_{2}\right) \cdot
\left( \overrightarrow{s}_{1}\times \overrightarrow{s}_{2}\right) -\left(
\overrightarrow{r}_{1}\times \overrightarrow{s}_{2}-\overrightarrow{r}%
_{2}\times \overrightarrow{s}_{1}\right) ^{2}=\left( \overrightarrow{r}%
_{1}\cdot \overrightarrow{s}_{2}-\overrightarrow{r}_{2}\cdot \overrightarrow{%
s}_{1}\right) ^{2},
\end{gather}
which, through definitions (\ref{spinors}) and (\ref{spinorss-2}), matches
Eq. (\ref{resss}).

\section{\label{Proof-Completeness}A Complete Basis for $SO\left( n,\mathbb{C%
}\right) $-invariant Polynomials}

We now proceed to prove that every $SO\left( n,\mathbb{C}\right)
$-invariant \textit{polynomial} $\mathcal{P}\left( \mathbf{A}\right)
$ of a $n\times n$
complex symmetric matrix $\mathbf{A}=\mathbf{A}^{T}\in M_{n}\left( \mathbb{C}%
\right) $ is a polynomial of\footnote{%
For $p>n$, Tr$\left( \mathbf{A}^{p}\right) $ can be expressed in terms of $%
\left\{ Tr\left( \mathbf{A}^{p}\right) \right\} _{1\leqslant
p\leqslant n}$ since, by virtue of the Cayley-Hamilton Theorem,
$\mathbf{A}$ fulfills its characteristic equation det$\left( \lambda
\mathbb{I}_{n}-\mathbf{A}\right)
=0$ ($\mathbb{I}_{n}$ denoting the $n\times n$ identity).} $\left\{ \text{Tr}%
\left( \mathbf{A}^{p}\right) \right\} _{1\leqslant p\leqslant n}$ :
\begin{equation}
\forall g\in SO\left( n,\mathbb{C}\right) ,~\mathcal{P}\left( g^{-1}\mathbf{A%
}g\right) =\mathcal{P}\left( \mathbf{A}\right) \Longrightarrow \mathcal{P}%
\left( \mathbf{A}\right) =\mathcal{Q}\left( \text{Tr}\mathbf{A},\text{Tr}%
\left( \mathbf{A}^{2}\right) ,...,\text{Tr}\left(
\mathbf{A}^{n}\right) \right) ,  \label{complex}
\end{equation}
where $\mathcal{Q}$ denotes some polynomial.

In order to prove (\ref{complex}), we start by observing that $\mathcal{P}%
\left( g^{-1}\mathbf{A}g\right) $ (which is holomorphic in $g$ and $\mathbf{A%
}$) is determined by analytic continuation\footnote{%
We thank Prof. Michel Dubois Violette for this argument.} from its
value
taken for $g\in SO\left( n,\mathbb{R}\right) $ and for $\mathbf{A}=\mathbf{A}%
^{T}\in GL\left( n,\mathbb{R}\right) $ (\textit{e.g.} through a
convergent series in the neighbourhood of any real point in
$\mathbb{C}$).

By virtue of this observation, it then suffices to prove
(\ref{complex}) for $g\in SO\left( n,\mathbb{R}\right) $ and for
$\mathbf{A}=\mathbf{A}^{T}\in M_{n}\left( \mathbb{R}\right) $.

In order to do so, we notice that $\mathbf{A}$ real symmetric can
always be
diagonalised through a suitable transformation $t\in SO\left( n,\mathbb{R}%
\right) $, yielding real eigenvalues $\left\{ \lambda
_{1},...,\lambda _{n}\right\} $. Thus, every $SO\left(
n,\mathbb{R}\right) $-invariant \textit{polynomial}
$\mathcal{P}\left( \mathbf{A}\right) $ of a real symmetric $n\times
n$ matrix $\mathbf{A}$ is also a polynomial in $\left\{
\lambda _{1},...,\lambda _{n}\right\} $; its $SO\left( n,\mathbb{R}\right) $%
-invariance implies\footnote{%
Indeed, an even permutation of the indices $\left\{ 1,...,n\right\}
$ can be represented by a suitable transformation of $SO\left(
n,\mathbb{R}\right) $.} that it is symmetric under even permutations
of the indices $\left\{ 1,...,n\right\} $. Furthermore, such a
polynomial in $\left\{ \lambda _{1},...,\lambda _{n}\right\} $ can
be split into a symmetric component and
into an antisymmetric component under odd permutations of the indices $%
\left\{ 1,...,n\right\} $:

\begin{itemize}
\item  the \textit{symmetric} component is given by $\frak{P}\left( \lambda
_{1},...,\lambda _{n}\right) $, a polynomial which is
\textit{symmetric} under \textit{all} permutations of indices
$\left\{ \lambda _{1},...,\lambda _{n}\right\} $. Its functional
dependence on the eigenvalues $\left\{ \lambda _{1},...,\lambda
_{n}\right\} $ can be proved to be as follows (see \textit{e.g.}
\cite{McDonald}):
\begin{equation}
\frak{P}\left( \lambda _{1},...,\lambda _{n}\right) =\widetilde{\frak{P}}%
\left( \sigma _{p}\equiv \sum_{i}\lambda _{i}^{p},~1\leqslant
p\leqslant
n\right) =\mathcal{Q}\left( \text{Tr}\mathbf{A},\text{Tr}\left( \mathbf{A}%
^{2}\right) ,...,\text{Tr}\left( \mathbf{A}^{n}\right) \right) .
\label{symmetric}
\end{equation}

\item  the \textit{anti-symmetric} component is of the form (recall (\ref
{symmetric}); see \textit{e.g.} \cite{McDonald})
\begin{equation}
\frak{P}^{\prime }\left( \lambda _{1},...,\lambda _{n}\right) \Delta
\left(
\lambda _{1},...,\lambda _{n}\right) =\mathcal{Q}^{\prime }\left( \text{Tr}%
\mathbf{A},\text{Tr}\left( \mathbf{A}^{2}\right)
,...,\text{Tr}\left( \mathbf{A}^{n}\right) \right) \Delta \left(
\lambda _{1},...,\lambda _{n}\right) ,  \label{antisymmetric}
\end{equation}
where
\begin{equation}
\Delta \left( \lambda _{1},...,\lambda _{n}\right) \equiv
\prod_{1\leqslant i<j\leqslant n}\left( \lambda _{j}-\lambda
_{i}\right) =\text{det}\left(
\begin{array}{cccc}
1 & \lambda _{1} & .... & \lambda _{1}^{n-1} \\
1 & \lambda _{2} & .... & \lambda _{2}^{n-1} \\
... & ... & .... & ... \\
1 & \lambda _{n} & .... & \lambda _{n}^{n-1}
\end{array}
\right)   \label{Vandermonde}
\end{equation}
is the Vandermonde determinant. It should be remarked that $\Delta $
is \textit{not} a polynomial in $\mathbf{A}$, because it is a square
root of
the discriminant of the characteristic polynomial det$\left( \lambda \mathbb{%
I}_{n}-\mathbf{A}\right) $. Indeed, in general the transformation
$t\in SO\left( n,\mathbb{R}\right) $ which diagonalizes $\mathbf{A}$
is not a polynomial in $\mathbf{A}$ itself, because, due to
orthonormalisation of the eigenvectors of $\mathbf{A}$, it involves
square roots.
\end{itemize}

Therefore, since we restrict to consider $SO\left( n,\mathbb{R}\right) $%
-invariant \textit{polynomials} in the real symmetric matrix
$\mathbf{A}$, the observations above lead to the conclusion that
such polynomials necessarily are of the form (\ref{symmetric}). The
analytic continuation of this result to $\mathbb{C}$ yields the
proof of (\ref{complex}).

Thus, as mentioned in Sec. \ref{SO_h(2,2)}, by means of
\textit{Newton's formul\ae } \cite{Newton-Refs} (\textit{cfr.} Eqs.
(\ref{a})-(\ref{d}) in
the case $n=4$), the characteristic polynomial det$\left( \lambda \mathbb{I}%
_{n}-\mathbf{A}\right) $ can be considered as a generating function
for the
ring of $SO\left( n,\mathbb{C}\right) $-invariant \textit{polynomials} of $%
\mathbf{A}=\mathbf{A}^{T}\in M_{n}\left( \mathbb{C}\right) $.
$\blacksquare $

\end{document}